\documentclass[aps,showpacs,floatfix,superscriptaddress,twocolumn]{revtex4-1}
\usepackage{graphicx}
\usepackage{amssymb}
\usepackage{graphics, color}
\usepackage{amsmath}
\usepackage{overpic}
\usepackage{float}
\usepackage{hyperref}
\usepackage{enumitem}
\newcommand{\wid}{0.45}

\newcommand*\pFq[2]{\;{}_{#1}F_{#2}}
\newcommand{\dos}{\nu(0)}
\usepackage{amsmath}

\DeclareMathOperator{\sech}{sech}

\newcommand{\dd}{d}
\graphicspath{{pics/}}

\definecolor{Gray}{gray}{0.4}

\begin{document}
\title{Global critical temperature in inhomogeneous superconductors induced by multifractality}
\author{James Mayoh}

\author{Antonio M. Garc\'\i a-Garc\'\i a}
\affiliation{TCM Group, Cavendish Laboratory, University of Cambridge, JJ Thomson Avenue, Cambridge, CB3 0HE, UK}

\begin{abstract}
There is growing evidence, from experiments and numerical simulations, that a key feature of sufficiently disordered superconductors is the spatial inhomogeneity of the order parameter. However not much is known analytically about the details of its spatial distribution or the associated global critical temperature that signals the breaking of long-range order.
Here we address this problem for disordered systems around an Anderson transition characterized by multifractal one-body eigenstates. In the limit of weak multifractality and for weakly coupled superconductors we compute the superconducting order parameter analytically, including its energy dependence and statistical distribution in space. The spatial distribution of the order parameter is found to be always log-normal. The global critical temperature, computed by percolation techniques and neglecting phase fluctuations, is enhanced with respect to the clean limit only for very weakly coupled superconductors. Some enhancement still persists even in the presence of moderate phase fluctuations crudely modelled by increasing the percolation threshold. Our results are also consistent with experiments, where enhancement of the critical temperature is observed in Al thin films, a very weakly coupled metallic superconductor, but not in more strongly coupled materials. 
\end{abstract}
\date{\today}

\maketitle

For many years the role of disorder in superconductivity was believed to be well understood. According to the so called Anderson theorem 
\cite{Anderson1959}, also stated independently by Gor'kov and Abrikosov \cite{Abrikosov1961}, the critical temperature of a conventional weakly-coupled superconductor is not affected by weak non-magnetic impurity scattering. These results are based on the assumption that
the local density of states in the material is unaffected by weak disorder \cite{Kim1993,Abrikosov1994}.
 However with the development of the Bogoliubov-de Gennes theory of superconductivity \cite{DeGennes1964} it became clear that the order 
parameter becomes increasingly inhomogeneous with increasing disorder. 

Experimentally it is well established
 \cite{Haviland1989,goldman1993,Nishida1982,Furubayashi1985,Alekseevskii1983,Bishop1985,Graybeal1985,Driessen2012,Tashiro2008}, especially for 
conventional superconducting thin films, that the critical temperature decreases monotonically as disorder increases. 
 Analytic results \cite{Maekawa1982,Maekawa1984}, obtained using mesoscopic techniques, confirmed that the interplay between weak disorder 
and Coulomb interactions could explain this suppression of the critical temperature. For stronger disorder around the superconductor insulator 
transition there is recent numerical \cite{Ghosal2001,Bouadim2011} evidence that, even in the absence of Coulomb interactions, phase 
fluctuations are enhanced \cite{Mondal2011} and the superconducting order parameter becomes highly inhomogeneous \cite{Trivedi2012,Sherman2014}.  Close to the Berezinski-Kosterlitz-Thouless transition phase correlation only persist along a ramified network, reminiscent of a percolation transition \cite{erez2013}. This is consistent with experimental observations  of a universal scaling of the order 
parameter amplitude distribution function\cite{Lemarie2013}, emergent granularity \cite{brun2014,noat2013} and reports of glassy features\cite{Ioffe2010}, with a supercurrent flow pattern reminiscent of a 
percolative cluster\cite{Seibold2012}, a pseudo-gap phase \cite{Bouadim2011,Mondal2013} and preformed Cooper pairs \cite{Sacepe2011} for 
sufficiently strong disorder. 

The upshot of this discussion is that the order parameter in the presence of strong disorder is highly inhomogeneous with strong phase fluctuations which makes it unlikely that superconductivity can be more robust than in the clean limit. The Anderson theorem does not really apply in this region as self-averaging, one of its assumptions, is not expected to hold for sufficiently strong disorder \cite{suslov2013}.   
However, recent theoretical studies have suggested that enhancement might indeed occur in the presence of strong disorder. The density matrix renormalization group analysis of ref. \cite{Tezuka2010} showed that phase coherence in a one dimensional disordered Hubbard model with attractive interactions at zero temperature is enhanced for weak coupling and disorder close to but below the superconductor-insulator threshold. In Refs. \cite{Feigelman2007,Burmistrov2012,Feigelman2010} it was reported that superconductivity was strongly enhanced around the Anderson metal-insulator transition.
The origin of this enhancement is directly related to the multifractality of eigenstates of the one-body problem in the critical regime
\cite{Wegner1980,Castellani1986,Falko1995}. The strong spatial correlations of multifractal eigenstates \cite{Fyodorov1997} around the Fermi energy lead to a more robust superconducting state as a consequence of two facts: the critical temperature, defined in these works as the temperature for which the order parameter at the Fermi energy vanishes, depends as a power-law, instead as an exponential, on the electron-phonon coupling constant. Moreover it is proportional to $E_0 \gg \epsilon_D$ and not to the Debye energy $\epsilon_D$ as is the case for conventional superconductors with no disorder. The energy scale 
$E_0$ is a cut-off related to the minimum length scale for which the eigenfunctions are multifractal. 
 At the Anderson transition this length scale is of the order of the mean free path and $E_0$ is of the order of the Fermi energy. 
However this critical temperature cannot be the maximum temperature at which a supercurrent is observed since it would lead to completely unrealistic critical temperatures of the order of the Fermi temperature of the material. This is not surprising as the analysis of refs.\cite{Burmistrov2012,Feigelman2007,Feigelman2010} does not take into account effects such as the spatial inhomogeneity of the superconductor, a key ingredient to understand the physics around the transition. Despite these limitations, the proposal that multifractality might have a profound impact on superconductivity is intriguing and deserves further investigation. 


In this paper we revisit the problem of a disordered weakly coupled superconductor in the limit of 
weak multifractality and including explicitly the effect of spatial inhomogeneities of the order parameter. In this region the effect of disorder is relevant but it is still possible to obtain explicit analytical results as a mean-field Bardeen-Cooper-Schrieffer (BCS) approach is still 
qualitatively valid. Weak multifractality is relevant in a variety of problems: two dimensional weakly disordered superconductors for system sizes much smaller than the localization length \cite{Falko1995}, weakly disordered $2+\epsilon$ superconductors in the $\epsilon \ll 1$ limit \cite{Wegner1980}, two dimensional disordered superconductors with spin-orbit interactions \cite{Hikami1980} and one dimensional superconductors with long range hopping \cite{Lobos2013}. 

The main conclusions of our study are as follows:
\begin{enumerate}[label=(\alph*)]
\item the spatial distribution function of the order parameter, and the associated local critical temperature, is always log-normal.

\item the global critical temperature of the sample, defined as the maximum temperature at which a supercurrent can flow, resulting from a percolation analysis, is very sensitive to the strength of the electron-phonon coupling constant. In all cases the global critical temperature is substantially lower than for a homogeneous order parameter computed at the Fermi energy. We only find an enhancement of this critical temperature, with respect to the bulk non-disordered limit, for very weak electron-phonon coupling.
 
 
 \item a crude estimation of the effect of phase fluctuations, induced by the Coulomb interaction or other processes, that suppresses superconductivity shows that in a realistic situation a substantial enhancement of the global critical temperature by disorder might be possible only in very weakly coupled materials such as aluminium. This is in qualitative agreement with the experimental observations of enhancement of the critical temperature in Al thin film \cite{Abeles1966,goldman1993}, but not in other more strongly coupled materials,  in a region of parameters for which multifractality might be relevant. 
\end{enumerate}

The paper is organized as follows. We first derive exact expressions for the superconducting gap and the critical temperature at the Fermi energy and its leading energy dependence as a function of the multifractal exponents. These exponents are directly related to the conductance of the material. 
Next we calculate analytically the full statistical distribution of the order parameter and the critical temperature in real space. The distribution is always log-normal and shows a highly inhomogeneous pattern with emergent granularity as disorder increases. We then compute the global critical temperature by assuming that the transition is induced by percolation.
A rough estimation of the suppression of the global critical temperature due to phase fluctuations is then carried out by slightly increasing the percolation threshold. Finally we discuss the limitations of the model and the relevance of our results for experiments. 

\section{BCS superconductivity and multifractality}

The natural framework to study the interplay of superconductivity and disorder is that of the Bogouliubov-de Gennes(BdG) theory of superconductivity \cite{DeGennes1964,DeGennes1966}. In this formalism an inhomogeneous mean-field BCS Hamiltonian,
\begin{equation}
\begin{split}
H =& \int \dd {\bf r} \left[\sum_\sigma \Psi^\dagger_\sigma({\bf r}) \left( - \frac{\hbar^2}{2m}\nabla^2 +U({\bf r}) -\mu \right)\Psi_\sigma({\bf r}) \right. \\ & \left. \vphantom{ \left( - \frac{\hbar^2}{2m}\nabla^2\right)} + \Delta({\bf r})\Psi_\uparrow^\dagger({\bf r})\Psi_\uparrow^\dagger({\bf r}) + {\mathrm{ h.c.} }\right]
\end{split}
\end{equation}
where $\Psi_\sigma^\dag({\bf r})$ creates an electron in position eigenstate ${\bf r}$ and spin $\sigma$ and $U({\bf r})$ is the random potential,
is diagonalized by the generalized Bogoliubov transformation, 
\begin{equation}
\begin{split}
\Psi_\uparrow({\bf r})=\sum_{\bf n}\left( u_{\bf n}({\bf r})\gamma_{\uparrow,{\bf n}} - v_{\bf n}^*({\bf r})\gamma_{\downarrow,n}^\dagger\right)\\
\Psi_\downarrow({\bf r})=\sum_{\bf n} \left( u_{\bf n}({\bf r})\gamma_{\downarrow,{\bf n}} + v^*_{\bf n}({\bf r})\gamma_{\uparrow,{\bf n}}^\dag\right)
\end{split}
\end{equation}
 where the coherence factors $v_{\bf n}({\bf r})$ and $u_{\bf n}({\bf r})$ depend on the index ${\bf n}$ that labels some convenient basis set for the problem. The superconducting state is characterized by the space dependent order parameter $\Delta({\bf r})$,
\begin{equation}
\Delta({\bf r})=-\frac{\lambda}{\dos}\langle\Psi_\uparrow({\bf r})\Psi_\downarrow({\bf r})\rangle.
\end{equation}
where $\lambda$ is the dimensionless BCS coupling constant and $\dos$ is the bulk density of states at the Fermi energy.
One drawback of this approach is that the resulting BdG equations can only be solved numerically. However, it has recently \cite{Shanenko2008} been shown that in the weak coupling limit and when the spatial inhomogeneities are not very strong, it may be assumed that $u_{\bf n}({\bf r}),v_{\bf n}({\bf r})$ are proportional to the eigenstates of the one-body problem $\psi_{\bf n}({\bf r})$. It is then straightforward to show that the BdG equations turn into a modified BCS gap equation,

\begin{equation}\label{gap_eq}
\Delta(\epsilon)=\frac{\lambda}{2}\int_{-\epsilon_D}^{\epsilon_D}\frac{I(\epsilon,\epsilon')\Delta(\epsilon')}{\sqrt{\epsilon'^2+\Delta^2(\epsilon')}}\tanh\left(\frac{\beta\sqrt{\epsilon'^2+\Delta^2(\epsilon')}}{2}\right)\dd\epsilon'
\end{equation}
where $\epsilon_D$ is the Debye energy which gives the energetic cutoff for the electron-phonon coupling, $\Delta(\epsilon)$ is the superconducting gap as a function of energy, $\beta=(k_BT)^{-1}$ with $T$ the system temperature, $I(\epsilon,\epsilon')=V\int\dd{\bf r}|\psi(\epsilon,{\bf r})|^2|\psi(\epsilon',{\bf r})|^2$ are the BCS interaction matrix elements and $\psi(\epsilon,{\bf r})$ is the eigenstate of the one-body problem of energy $\epsilon$. An identical result is obtained from a generalized BCS variational approach. In both cases the spatial dependence of the gap 
\cite{Ma1985, Ghosal2001} is given by,
\begin{equation}\label{twf0}
\Delta({\bf r})=\frac{\lambda V}{2}\int \frac{\Delta(\epsilon)}{\sqrt{\Delta(\epsilon)^2+\epsilon^2}}|\psi(\epsilon,{\bf r})|^{2}d\epsilon.
\end{equation}
This model, already employed in the literature of disordered superconductors \cite{Feigelman2007}, has several appealing features. By using supersymmetric \cite{efetov1983supersymmetry,Falko1995}, and other non-perturbative techniques, explicit analytic expressions for the matrix elements $I(\epsilon,\epsilon')$ 
for a broad range of disorder strengths
 can be found \cite{Mirlin2000}. It is also well established that for disordered systems close to the metal to insulator transition the eigenfunctions are multifractal \cite{Mirlin2000,Evers2008}. A commonly used measure for multifractality is the anomalous scaling of the inverse participation ratio (IPR) \cite{Wegner1980,Castellani1986},
\begin{equation}
P_q=\int\dd{\bf r}|\psi({\bf r})|^{2q}\sim L^{d_q(q-1)},
\end{equation}
where $d_q < d$ is a multifractal dimension. This can be extended to the slow energy decay of eigenfunction correlations at different energies, 
\begin{equation}\label{ME}
I(\epsilon,\epsilon')=\left(\frac{E_0}{|\epsilon-\epsilon'|}\right)^\gamma
\end{equation}
so long as $\delta_L\ll |\epsilon-\epsilon'|<E_0$, where $\gamma=1-\frac{d_2}{d}$. 
The energy scale $E_0=(\dos L_0^3)^{-1}$ is associated with the large energy cutoff in fractal behaviour and $L_0$ is the short length scale cutoff associated with fractal behaviour. Below the metal-insulator transition it is expected $L_0$ should be of similar size to the mean free path, $\ell$. The Ioffe-Regel criterion $k_F \ell\sim1$ implies that at the mobility edge $E_0\sim E_F$. In systems with weaker disorder $E_0 \ll E_F$ but typically, at least for weakly coupled  metallic superconductors, it is still much larger than other energy scales such as the Debye energy or the superconducting gap.

The parameter $0\leq\gamma\leq1$ describes the strength of multifractality in the system. In particular the scaling exponents $d_q$ depend on the specific model chosen and the degree of disorder. As we mentioned previously our formalism is only valid in the limit of weak coupling and not very strong spatial inhomogeneities. Weak multifractality, $\gamma \ll 1$, can still occur in this limit, for instance in weakly disordered metals in $2+\epsilon$ dimensions or in strictly two dimensions for sizes much smaller than the localization length. The full set of multifractal dimension in this case is known analytically \cite{Falko1995}, $d_q \approx d(1-\kappa q)$ with $\kappa = \alpha/g$, $g$ the dimensionless conductance and $\alpha = 1/2,(1)$ for systems with (broken) time-reversal invariance. We note that for sufficiently large $q$ deviations from this simple linear behaviour are expected but these corrections are in general negligible for the observables of interest.
The limit $\gamma=0$ corresponds to zero disorder where the bulk metal behaviour is recovered, $I(\epsilon,\epsilon')=1$ leading to the usual expressions for the BCS gap,
$
\Delta_0 \approx 2\epsilon_D e^{-\frac{1}{\lambda}}$
 and the critical temperature, 
$
T_{c0} \approx \frac{2 e^{\gamma_E}}{\pi} \epsilon_D e^{-\frac{1}{\lambda}}
$
where $\gamma_E$ is the Euler-Mascheroni constant.

We have included explicitly in the gap equation the cut-off, the Debye energy $\epsilon_D$, related to the phonon coupling. This becomes particularly important in the limit $\gamma\to0$ as 
the BCS gap equation 
does not converge for $\epsilon_D\to\infty$. In the limit of weak multifractality, $\gamma \ll 1$, the gap equation is well defined for $\epsilon_D\to\infty$ but we shall see that in order to get meaningful results it is necessary to keep the physical cut-off $\epsilon_D$ finite. For $\gamma\approx 1$ it is plausible that the effective cutoff induced by the matrix elements will make $\epsilon_D$ less important \cite{Feigelman2007}. However in this limit the approximation $\Delta\lesssim\delta_L$ breaks down and the BCS mean-field theory is no longer valid. It should also be noted that the matrix element, Eq. (\ref{ME}), neglects contributions from the region $|\epsilon-\epsilon'|\sim\delta_L$ where $\delta_L$ is the mean level spacing, which will become increasingly important in the case of strong fractality. We show in appendix \ref{Ap:dl} that neglecting the effect of $\delta_L$ is valid in the limit of weak multifractality $\gamma\ll1$, $\delta_L \ll \epsilon_D$ we are interested in. 
\section{Energy dependence of the order parameter at zero temperature}
\begin{figure}[h!]
\begin{centering}
\includegraphics[width=\wid\textwidth]{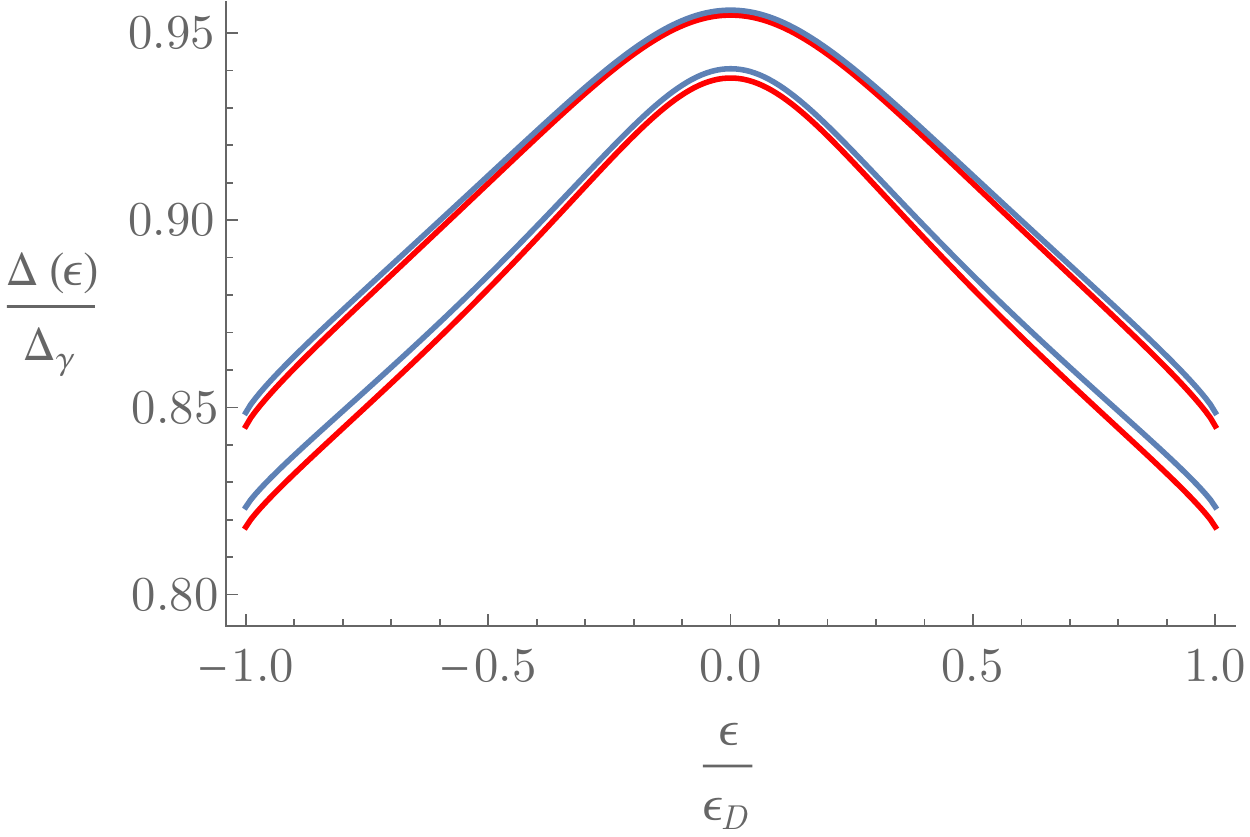}
\includegraphics[width=\wid\textwidth]{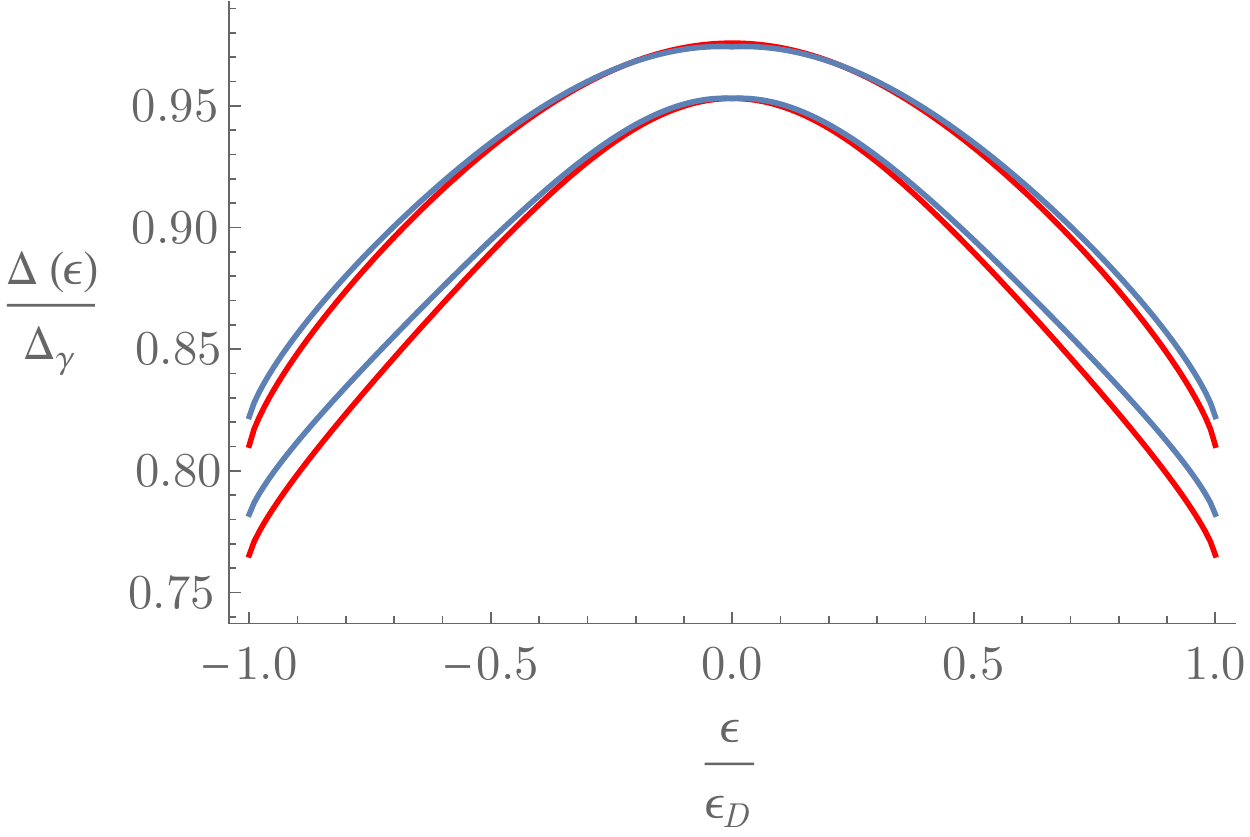}
\caption{Energy dependence of the gap $\Delta(\epsilon)$. Comparison between the numeric results from Eq.(\ref{gap_eq_t0}) (red) and the analytical calculation $\Delta(\epsilon)=\Delta_\gamma(1+\gamma f_1(\epsilon)+\gamma^2f_2(\epsilon))$ (blue) from Eqs. (\ref{f1_wrk}), (\ref{f2}) with $\lambda=0.3$ and $\gamma=0.1$ ({\it Upper Plot}) and
 $\gamma=0.2$ ({\it Lower Plot}). In both cases the upper pair of lines correspond to $E_0/\epsilon_D=100$ and the lower pair of lines to $E_0/\epsilon_D=20$. 
 We observe an excellent agreement in the full range of energy. The decay depends only on the degree of multifractality.
}\label{F_num_gap}
\end{centering}
\end{figure}

As a first step to compute analytically, in the limit of weak multifractality $\gamma\ll1$, the spatial distribution of the order parameter we solve the gap equation at zero temperature,
\begin{equation}\label{gap_eq_t0}
\Delta(\epsilon)=\frac{\lambda}{2}\int_{-\epsilon_D}^{\epsilon_D}\frac{\Delta(\epsilon')}{\sqrt{\epsilon'^2+\Delta^2(\epsilon')}}\left|\frac{E_0}{\epsilon-\epsilon'}\right|^\gamma d\epsilon'
\end{equation}
including its energy dependence. 
We expand the left-most parts of the gap equation in powers of $\gamma$ using the ansatz,
\begin{equation}\label{gapenergy}
\Delta(\epsilon)=\Delta_\gamma(1+\gamma f_1(\epsilon)+\gamma^2f_2(\epsilon) +\ldots).
\end{equation}
By using standard techniques, detailed in appendix \ref{Ap_S2}, we obtain results for $\Delta_\gamma,f_1(\epsilon),f_2(\epsilon)$. The expansion may be easily continued to arbitrarily high order 
however for weak multifractality this is clearly unnecessary. The explicit, but rather cumbersome, analytical expressions for $f_1(\epsilon),f_2(\epsilon)$ Eqs.(\ref{f1_wrk}),(\ref{f2}),  to be found in the appendix \ref{Ap_S2}, are in very good agreement, figure \ref{F_num_gap}, with the numerical solution of Eq.(\ref{gap_eq_t0}). We refer to the appendix \ref{Ap:num} for more details on the numerical calculation.

Several comments are in order: a) the energy dependence of the gap decays smoothly from the Fermi energy with an exponent that depends only on $\gamma$,
b) $h_1(\epsilon), h_2(\epsilon)$ are such that $h_i(0)=0$ and $h_i(\epsilon)$ is an even function in $\epsilon$. This means that $\Delta(0)=\Delta_\gamma(1+\gamma c_1+\gamma^2c_2 )$. The leading correction $c_1<0$ is negative as the zeroth order $(E_0/|\epsilon|)^\gamma$ term of the expansion is an overestimation of the exact matrix elements Eq.(\ref{ME}). Increasing $E_0$ results in smaller $c_i$ and thus the peak of $\Delta(\epsilon)$ is closer to $\Delta_\gamma$,
c) unsurprisingly, increasing $\gamma$ results in a larger error in the analytic results and in a greater difference between the peak value $\Delta(0)$ and the minima $\Delta(\pm\epsilon_D)$.
\subsection{$\Delta_\gamma$ and the associated critical temperature $T_{c\gamma}$}\label{sec:zero_beh}
The gap $\Delta_\gamma$ in Eq.(\ref{gapenergy}) is defined as the maximum of the order parameter $\Delta(\epsilon)$ in a disordered system characterized by weak multifractality. It corresponds approximately its value at the Fermi energy. 
An interesting question to consider in the later study of spatial inhomogeneities and enhancement of superconductivity is how $\Delta_\gamma$ differs from the its value in the clean limit, $\Delta_0$. 
 
An exact analytical expression of  $\Delta_\gamma$  is available, see Eq.({\ref{cf1}}) of the supplementary information. However it is more illuminating to carry out an expansion of Eq.({\ref{cf1}}) about $\epsilon_D/\Delta_\gamma\to\infty$, a limit that always holds for weakly coupled superconductors and should therefore be valid for $\gamma\ll 1$. Expanding to first order and solving for $\Delta_\gamma$ we find,
\begin{equation}\label{gap1}
\Delta_\gamma=D(\gamma)\epsilon_D\left(1+\frac{\gamma}{\lambda}\left(\frac{\epsilon_D}{E_0}\right)^\gamma\right)^{-\frac{1}{\gamma}}
\end{equation}
where,
\begin{equation}\label{dgamma}
D(\gamma)=\left(\frac{\gamma\Gamma(\frac{1}{2}(1-\gamma))\Gamma(\frac{\gamma}{2})}{2\sqrt{\pi}}\right)^{\frac{1}{\gamma}}
\end{equation}
and $\Gamma(x)$ is the usual Gamma function. It should be noted that as $E_0\to\infty$ the gap $\Delta_\gamma$ is still proportional to $\epsilon_D$, 
not to $E_0$ as in \cite{Feigelman2007} where $\Delta_\gamma \sim E_0 \lambda^{1/\gamma}$. 
The reason for this disagreement is that we have kept the Debye energy $\epsilon_D$ finite in our calculation. We believe that this is necessary since typically $\epsilon_D \ll E_0$ so it is not consistent to take the Debye energy to infinity while keeping $E_0$ finite.
This is also necessary to recover the BCS result in the limit $\gamma\to0$, as Eq.(\ref{gap1}) does.

\begin{figure}[h!]
\begin{centering}
\includegraphics[width=\wid\textwidth]{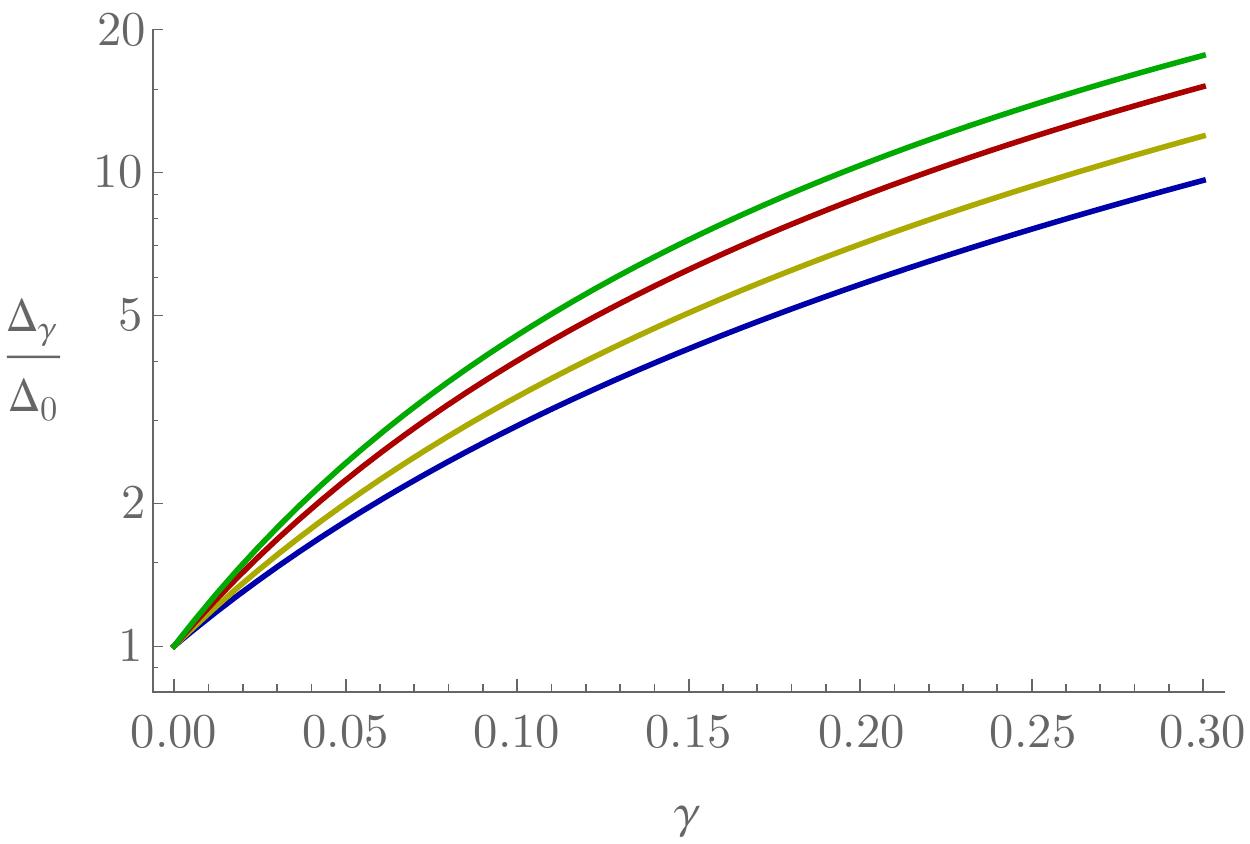}
\includegraphics[width=\wid\textwidth]{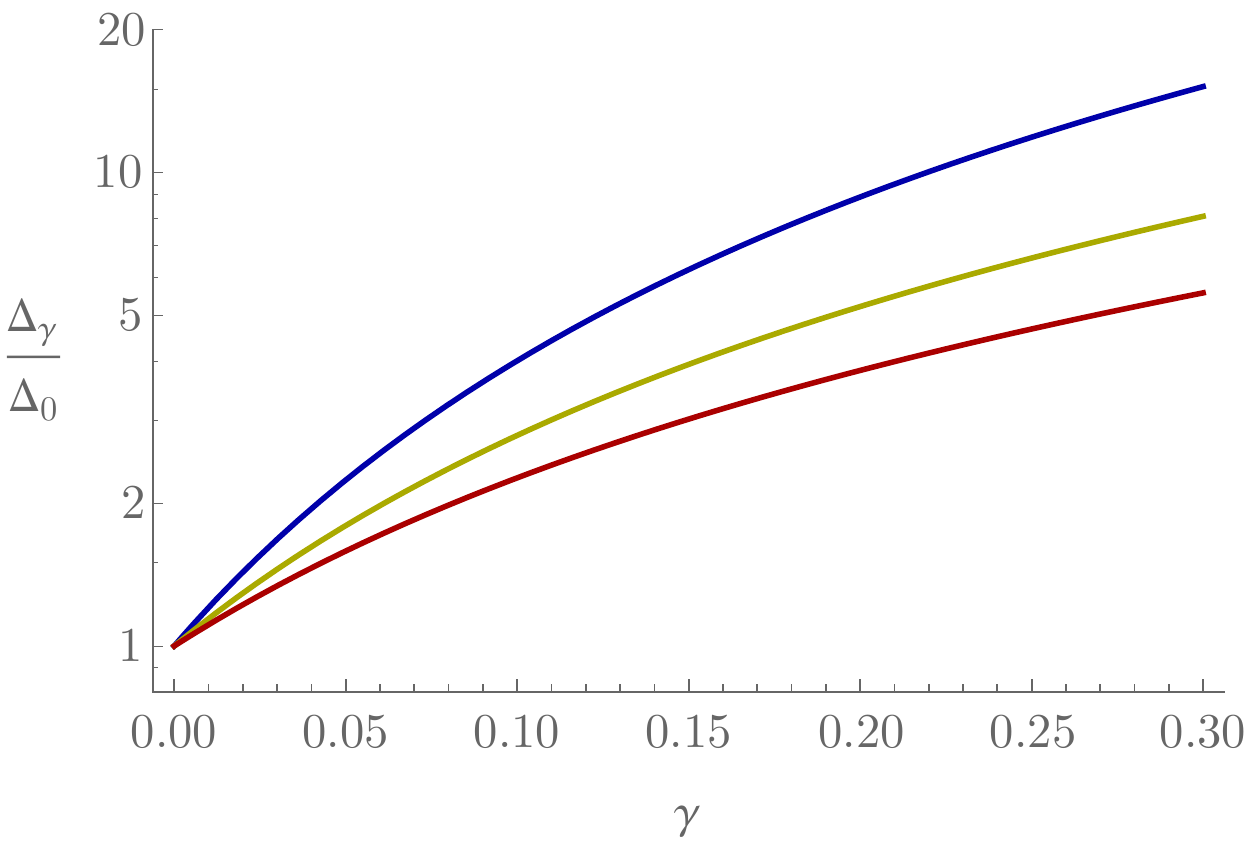}
\caption{{\it Upper:} The value of the gap at the Fermi energy $\Delta_\gamma$ from Eq.(\ref{cf1}) for $\lambda=0.3$ and $E_0/\epsilon_D=$10(Blue), 20(Yellow), 50(Red), 100(Green). {\it Lower:} $ E_0/\epsilon_D=50$ and $\lambda =$0.3(Blue), 0.4(Yellow), 0.5(Red). $\Delta_\gamma$ increases exponentially for $\gamma \approx 0$. The gradient decreases for larger $\gamma$. We show later that a large value of $\Delta_\gamma$ does not lead necesarily to a large enhancement of the critical temperature of the sample.}\label{F_gap}
\end{centering}
\end{figure}

In the limit of $\gamma\ll1$ we can re-express $\Delta_\gamma$ in the more transparent form,
\begin{equation}\label{gap_sg}
\Delta_\gamma\approx D(\gamma) \epsilon_D e^{-\frac{1}{\lambda}\left(\frac{\epsilon_D}{E_0}\right)^\gamma}
\end{equation}
with $D(\gamma)\approx2(1+\frac{\pi^2}{12}\gamma+\ldots)$. This result indicates that in the limit of weak fractality the gap behaves as if it has an effective coupling constant $\lambda_\text{eff}=\lambda\left(\frac{E_0}{\epsilon_D}\right)^\gamma$ giving rise to an exponential increase from $\Delta_0$ with increasing $\gamma$, see figure \ref{F_gap}. This is the reason why even a small value for $\gamma$, corresponding to weak disorder, can lead to substantial changes in the superconducting gap with respect to the clean limit provided that the effect of disorder is computed self consistently.
Another interesting parameter, 
 that describes a disordered system, is the temperature at which $\Delta(0)$ vanishes. This can be found by solving,
$1=\lambda\int_0^{\epsilon_D}\left(\frac{E_0}{\epsilon}\right)^\gamma\frac{\tanh(\beta_c\epsilon/2)}{\epsilon}\dd\epsilon$.
 This integration can also be carried out analytically, see appendix \ref{Ap:tc}, to give,
\begin{equation}\label{Tcg}
k_BT_{c\gamma}=\epsilon_D C(\gamma)\left(1+\frac{\gamma}{\lambda}\left(\frac{\epsilon_D}{E_0}\right)^\gamma\right)^{-\frac{1}{\gamma}}
\end{equation}
where,
\begin{equation}\label{cgamma}
C(\gamma)=\left[2\gamma(2^{\gamma+1}-1)\,\Gamma(-\gamma)\,\zeta(-\gamma)\right]^{\frac{1}{\gamma}}
\end{equation}
and $\zeta(x)$ is the Riemann zeta function. In the limit $\gamma\to0$ this expression recovers the BCS result. It should be noted that the derivation of this result is independent from the derivation for $\Delta_\gamma$.

The ratio of Eq.(\ref{gap1}) and Eq.(\ref{Tcg}), $2\Delta_\gamma/T_{c\gamma}$ is a useful indicator of the relevance of disorder,
\begin{equation}\label{ratio}
\frac{2\Delta_\gamma}{k_B T_{c\gamma}}=\frac{2D(\gamma)}{C(\gamma)}
=2\left(\frac{\Gamma(\frac{1}{2}(1-\gamma))\Gamma(\frac{\gamma}{2})}{4\sqrt{\pi}(2^{\gamma+1}-1)\,\Gamma(-\gamma)\,\zeta(-\gamma)}\right)^{\frac{1}{\gamma}}
\end{equation}
As in the non-disordered case this ratio is independent of the material constants but it is now a function of the strength of the multifractal exponent, $\gamma$. Expanding about $\gamma=0$ we find,
\begin{equation}\label{ratio_lin}
\frac{2\Delta_\gamma}{k_B T_{c\gamma}}=2\pi e^{-\gamma_E}(1+\frac{1}{2}(\gamma_E^2-\frac{\pi^2}{12}+2\ln^2(2)+2\gamma^{sj}_1)\gamma +\mathcal{O}(\gamma^2))
\end{equation}
where $\gamma^{sj}_n$ is the Stieltjes Gamma function. Note that the BCS result $2\Delta_0/T_{c0} =2\pi e^{-\gamma_E}$, is recovered in the limit $\gamma\rightarrow0$.
The above expression is still valid to relatively large $\gamma$ as the corrections from higher order terms in the gap and critical temperature are expected to cancel to a good approximation. Indeed Eq. (\ref{ratio}) agrees well with recent numerical results focused on the vicinity of $\gamma\to1$\cite{Feigelman2007},
$\frac{2D(\gamma=1)}{C(\gamma=1)}=4$. However we emphasise that in this limit the BCS mean-field approach is in principle not applicable so we refrain from extracting physical conclusions. We also note that deviations from the BCS value for the ratio of the gap and critical temperature have been observed experimentally\cite{Sacepe2008}. However the observable measured in experiment are not defined identically to the theoretical ones above so direct comparison is not trivial.
In summary, these results appear to indicate that the gap and critical temperature in a disordered material can be substantially different from that in the clean limit. 

The inherent inhomogeneity induced by disorder will play an important role so we expect that both quantities vary substantially in space and therefore we must envisage a procedure to estimate the critical temperature of the sample defined as the maximum temperature for which a supercurrent can flow. To explore these issues we begin by calculating the statistical distribution of the gap in space. 
\section{Distribution of the order parameter in real space}
 In a disordered material the gap in real space is intrinsically inhomogeneous however for a particular disorder strength it should have a well defined statistical distribution. As was mentioned in the introduction, this spatial distribution function of the order parameter is an outstanding open problem in the theory of superconductivity. In this section we compute analytically this distribution function for the case of weak multifractality of the one-body eigenstates. We leave the details of the calculation to appendix \ref{Ap_S3} and here only sketch the main steps. The starting point is the space dependent gap $\Delta({\bf r})$ Eq.(\ref{twf0}), resulting from the generalised trial wave function method mentioned in the introduction, and the energy dependence of the order parameter Eq.(\ref{gapenergy}) computed in the previous section. The moments of $\Delta({\bf r})$ are given by,
\begin{equation}
\langle\Delta^n({\bf r})\rangle=\int\dd{\bf r}\prod_{j=1}^n\left(\frac{\lambda V}{2}\int\frac{\Delta(\epsilon_{j})}{\sqrt{\Delta(\epsilon_{j})^2+\epsilon_{j}^2}}|\psi(\epsilon_{j},{\bf r})|^{2}d\epsilon_j \right). 
\end{equation}

\begin{figure}[h!]
\begin{centering}
\hspace*{1.2cm}
\begin{overpic}[width=0.4\textwidth]{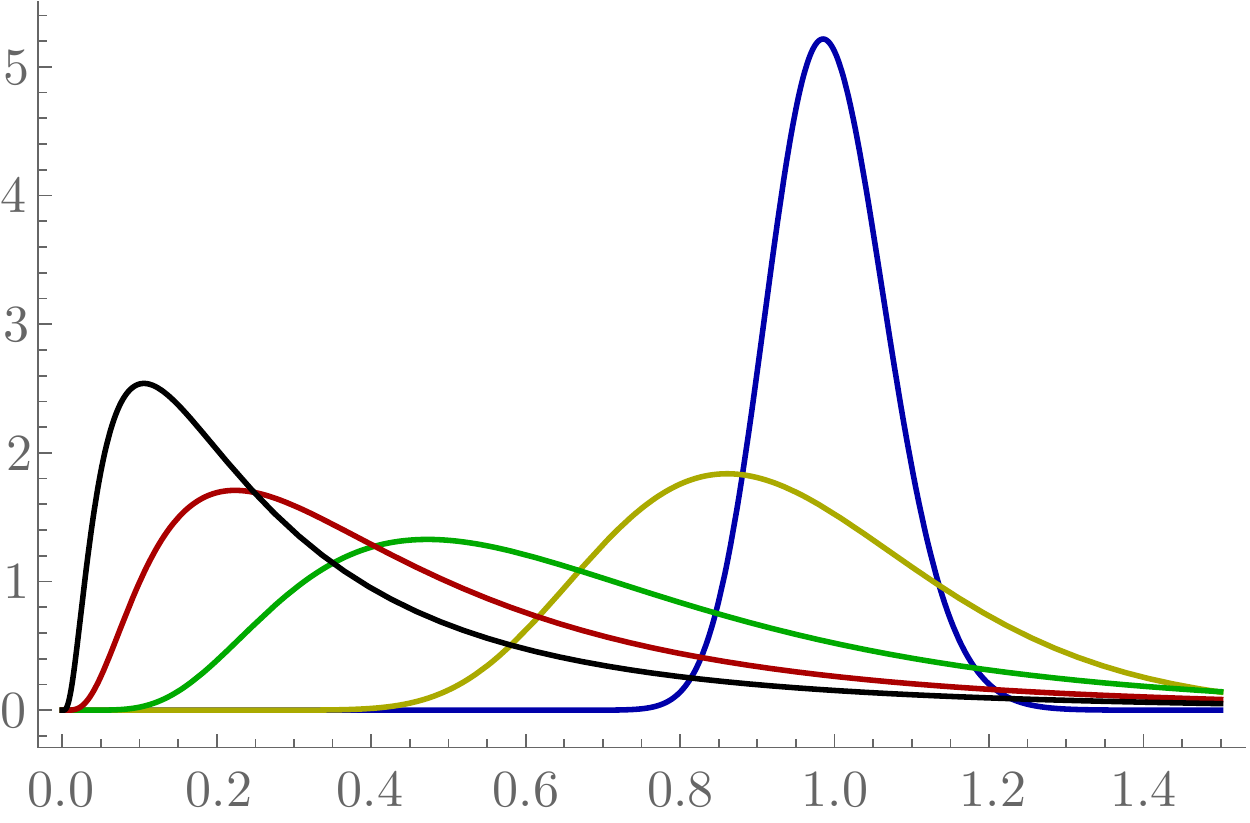}
\put(-18,35){\color{Gray} \small $\mathcal{P}\left(\frac{\Delta({\bf r})}{\Delta_\gamma}\right)$}
\put(45,-6){\color{Gray} \small $\frac{\Delta({\bf r})}{\Delta_\gamma}$}
\end{overpic}
\vspace*{0.5cm}

\hspace*{1.2cm}
\begin{overpic}[width=0.4\textwidth]{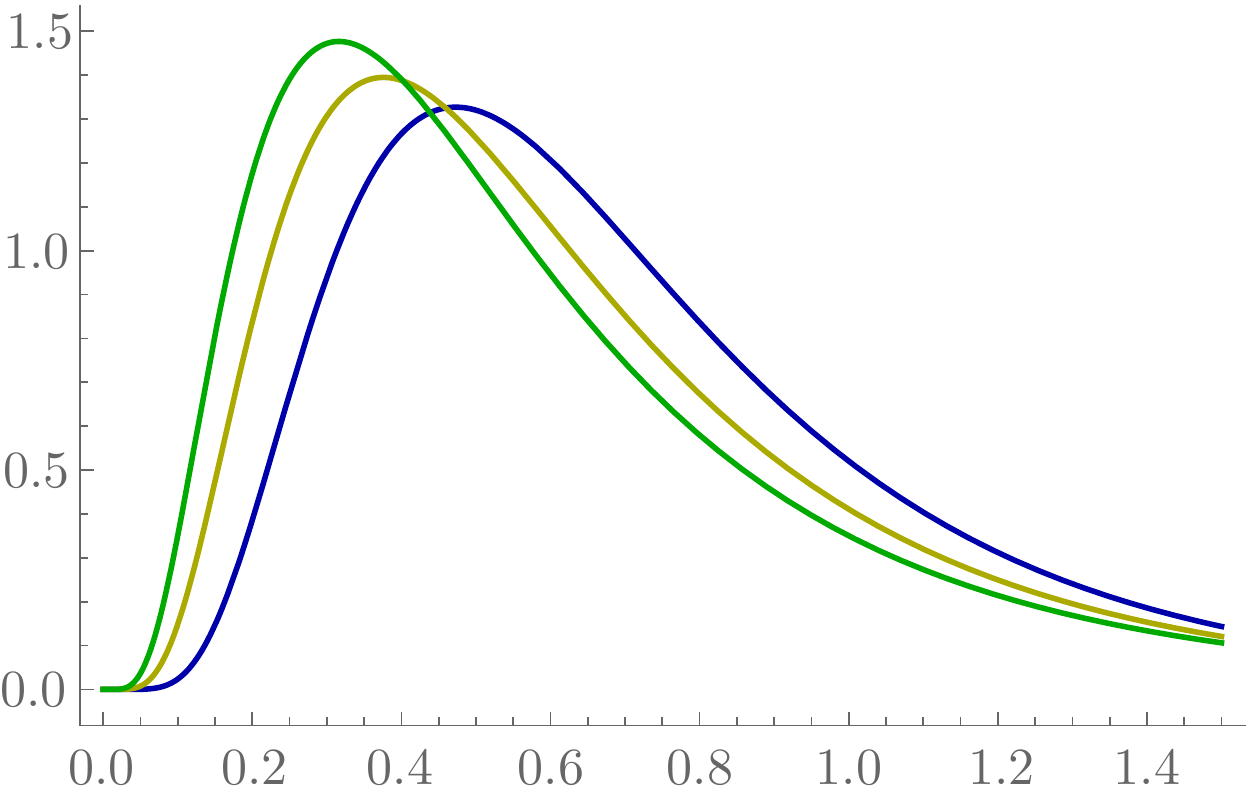}
\put(-18,35){\color{Gray} \small $\mathcal{P}\left(\frac{\Delta({\bf r})}{\Delta_\gamma}\right)$}
\put(45,-6){\color{Gray} \small $\frac{\Delta({\bf r})}{\Delta_\gamma}$}
\end{overpic}
\vspace*{0.5cm}
\caption{Probability distribution of the gap Eq.(\ref{PD}) for different choices of multifractality strength $\gamma = 2\kappa$ and $E_0$. {\it Upper}: $E_0/\epsilon_D=20$, $\kappa=$0.001(Blue), 0.01(Yellow), 0.05(Green), 0.1(Red), 0.15(Black) 
{\it Lower}: $\kappa=0.05$, $E_0/\epsilon_D$=20(Blue), 50(Yellow), 100(Green) where $\kappa^{-1}$ is proportional to the dimensionless conductance (see introduction). In the metallic limit $\kappa \to 0$ the distribution approaches a Dirac delta function centred on the value of the gap at the Fermi energy. For any finite $\kappa$ the distribution is log-normal. It becomes broader as $\kappa$ increases with a maximum that moves rapidly to smaller values of the gap. The distribution depends only weakly on $E_0$.}\label{F_pdr}
\end{centering}
\end{figure}

In the limit  $\gamma \ll 1$, and keeping only leading terms, it is possible to evaluate approximately the generalized eigenstate correlation function above and to compute explicitly the moments.
The final result is,
\begin{equation}
\frac{\langle\Delta^n({\bf r})\rangle}{\left(\Delta_\gamma\right)^{n}}
=e^{\kappa\ln(\epsilon_D/E_0) (3n-n^2)}
\end{equation}
where $\kappa$ is inversely proportional to the dimensionless conductance, $\gamma=2\kappa$, from which it is straightforward to show that the distribution function associated to these moments is log-normal,
\begin{equation}\label{PD}
\mathcal{P}\left(\frac{\Delta({\bf r})}{\Delta_\gamma}\right)=\frac{\Delta_\gamma}{\Delta({\bf r})\sqrt{2\pi}\sigma} \exp\left[-\frac{\left(\ln \left(\frac{\Delta({\bf r})}{\Delta_\gamma}\right)-\mu\right)^2}{2\sigma^2}\right]
\end{equation}
with $\mu=3\kappa\ln(\epsilon_D/E_0)$, $\sigma=\sqrt{2\kappa\ln(E_0/\epsilon_D)}$. The mean value for the distribution is,
\begin{equation}
\left\langle\frac{\Delta({\bf r})}{\Delta_\gamma}\right\rangle = \left(\frac{\epsilon_D}{E_0}\right)^{2\kappa}
\end{equation}
 and the variance is given by
\begin{equation}
\mathrm{Var}\left(\frac{\Delta({\bf r})}{\Delta_\gamma}\right)=\left(\frac{\epsilon_D}{E_0}\right)^{2\kappa}\left(1-\left(\frac{\epsilon_D}{E_0}\right)^{2\kappa}\right)
\end{equation}
As $E_0$ is typically large compared to $\epsilon_D$ the above results indicate that the mean value $\Delta({\bf r})$ can be much smaller than $\Delta_\gamma$ and also that the distribution may be rather broad. These values also indicate that the distribution of $\Delta({\bf r})$ is strongly affected by changes to the disorder strength, $\kappa$, but is rather weakly dependent on the value of $\epsilon_D/E_0$, see figure \ref{F_pdr}. This implies that the chosen value of $E_0$ and any dependence of $E_0$ on the disorder strength has little effect on our results provided that $\epsilon_D/E_0 \ll 1$.

\begin{figure*}
\begin{centering}
\includegraphics[width=0.85\textwidth]{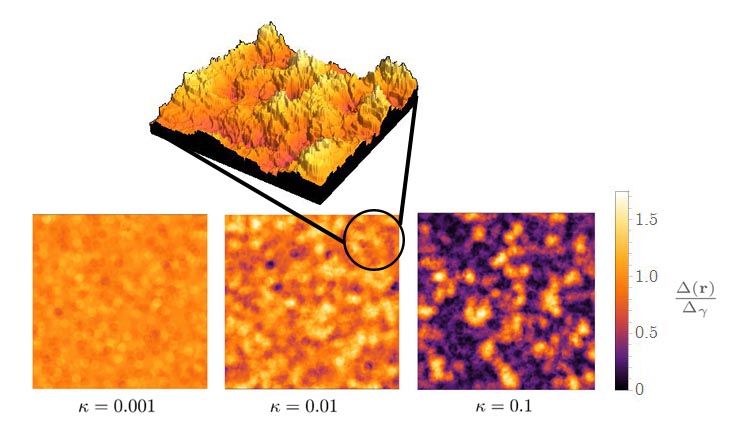}
\caption{The spatial dependence of the gap $\Delta({\bf r})/\Delta_\gamma$ obtained from the analytical prediction, a log-normal distribution Eq. (\ref{PD}), for three values of disorder, $\kappa = 0.001,0.01,0.1$ within the region of weak-multifractality $\kappa \ll 1$. As disorder increases, the regions for which enhancement is observed become increasingly sparse with large regions of very small values of the order parameter. In the upper plot we zoom a small spatial region for $\kappa=0.01$ in order to illustrate the intrincate spatial distribution of $\Delta({\bf r})/\Delta_\gamma$. This emergent granularity as disorder increases is qualitatively similar to that observed in recent numerical and experimental studies \cite{Lemarie2013,Sacepe2011,Ghosal2001} of disordered superconductors. We note that the local critical temperature has the same log-normal distribution Eq.(\ref{ratioexp}). It is therefore natural to estimate the global critical temperature of the sample by a percolation analysis.}\label{f_pp}
\end{centering}
\end{figure*}

In the limit $\kappa\to0$, 
\begin{equation}
\mathcal{P}\left(\frac{\Delta({\bf r})}{\Delta_\gamma}\right)=\delta\left(\frac{\Delta({\bf r})}{\Delta_\gamma}-1\right)
\end{equation}
this corresponds to the non-disordered case where the gap is uniform in space. Interestingly, as disorder increases, the distribution of the gap broadens and the mean moves to lower values with an extended tail up to $\Delta({\bf r})>\Delta_\gamma$, see figure \ref{F_pdr}. The decrease in the mean value follows physically from the confinement of the electrons to small regions when disorder is added. The gap is enhanced at some points of the material as the single electron wavefunctions are confined and overlap more strongly. However the reverse situation also occurs, resulting in many regions where the electron density and gap is reduced compared to the bulk. As disorder strength is increased and the degree of overlap in enhanced regions increases, the area of the suppressed regions also increases resulting in a decrease of the mean value of the distribution. This process is illustrated in figure \ref{f_pp}. It should be noted that the expansion to higher orders will modify and slightly broaden the distribution of Eq.(\ref{PD}). However our analytical result still provides a good approximation for the spatial distribution of the gap in the limit of weak multifractality. Indeed it is, see figure \ref{F_pdr}, qualitatively similar to that found in previous numerical and experiment studies \cite{Lemarie2013,Sacepe2011,Ghosal2001}.

\section{Distribution of $T_c({\bf r})$}
The inverse transformation of Eq. (\ref{twf0}) is given by, 
\begin{equation}\label{gwf_inv}
\Delta(\epsilon)=\int d{\bf r} \Delta({\bf r})|\psi(\epsilon,{\bf r})|^2
\end{equation}
In the case of finite temperature this should recover the gap equation Eq. (\ref{gap_eq}). This follows from the generalisation of the gap equation at finite temperature,
\begin{equation}\label{twf}
\begin{split}
&\Delta({\bf r})=\\ &\frac{\lambda V}{2}\int \frac{\Delta(\epsilon)}{\sqrt{\Delta(\epsilon)^2+\epsilon^2}}|\psi(\epsilon,{\bf r})|^{2}\tanh\left(\frac{\sqrt{\epsilon^2+\Delta^2(\epsilon)}}{2k_BT}\right)d\epsilon.
\end{split}
\end{equation}
It is clear solving for the critical temperature in equation $\Delta({\bf r})=0$ for all ${\bf r}$ will require that $T_c$ varies in space. We further know $k_B T_c(\epsilon)=\frac{C(\gamma)}{D(\gamma)}\Delta(\epsilon,T=0)$ solves the gap equation Eq. (\ref{gap_eq}) at $\Delta(\epsilon)\to0$ for all $\epsilon$. 
 It follows that the transformations which apply to the gap must also apply to the critical temperature,
\begin{equation}
T_c(\epsilon)=\int d{\bf r}T_c({\bf r})|\psi(\epsilon,{\bf r})|^2
\end{equation}
By comparison with Eq. (\ref{gwf_inv}),
\begin{equation}
 k_B T_c({\bf r})=\frac{C(\gamma)}{D(\gamma)}\Delta({\bf r},T=0)
\end{equation}
as one might have expected.
Whence the distribution function calculated for the gap in space will also hold for the critical temperature.
\begin{equation}\label{ratioexp}
\mathcal{P}\left(\frac{T_c({\bf r})}{T_{c_\gamma}}\right)=\mathcal{P}\left(\frac{\Delta({\bf r})}{\Delta_\gamma}\right)
\end{equation}
Next we employ this expression as the starting point to estimate the global critical temperature of the material by percolation techniques.

\section{Calculation of the global critical temperature of the sample using a percolation model}
The results from the previous section indicates that weak multifractality is responsible for the broad spatial distribution of the order parameter and the local critical temperature $T_c({\bf r_0})$. A natural question to ask is: what is the global critical temperature $T_c^{\text{mat}}$ of the material defined as the maximum temperature at which a supercurrent can flow? 
Recent work on inhomogeneous superconductors \cite{Mayoh2014a,erez2013} suggest that a percolation transition can be the driving force for the breakdown of phase coherence in an inhomogeneous system. Indeed many numerical studies have found that at strong disorder and finite temperature phase correlations become in general weakened due to the emergent granularity of the system \cite{Dubi2007,Ghosh2013}. More specifically long-range order is expected to be sustained by the persistence of phase correlations on a ramified network that permeates the system \cite{erez2013}.
 Certainly the critical temperature predicted by percolation of the amplitude of the order parameter is an upper bound for the global critical temperature $T_c^{\text{mat}}$ as phase fluctuations can still break long range order even if there exists a percolating cluster for the supercurrent to flow.

\begin{figure}[h!]
\begin{centering}
\includegraphics[width=\wid\textwidth]{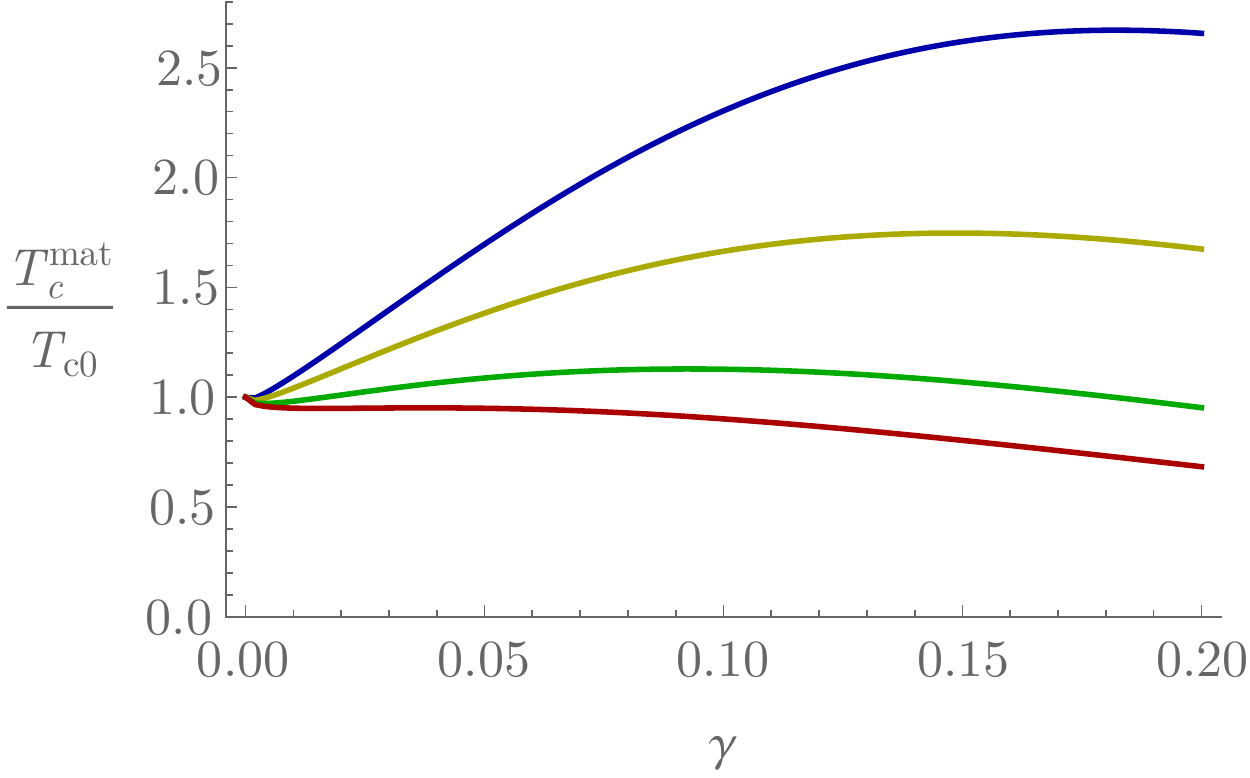}
\caption{The global critical temperature $T_c^{\text{mat}}$, from Eqs.(\ref{PD}) and (\ref{tcper}), obtained as the temperature at which the percolation transition occurs, $\phi_c=0.676$, in units of the BCS non-disordered critical temperature as a function of the degree of multifractality $\gamma$ for $E_0/\epsilon_D=100$ and $\lambda=0.25$(Blue), $0.3$(Yellow), $0.4$(Green), $0.5$(Red). Except in the case of small $\lambda$, no or very modest enhancement of $T_c^{\text{mat}}$ is observed as $\gamma$ increases. In all cases $T_c^{\text{mat}}$ moves well below $T_{c\gamma}$ Eq.(\ref{Tcg}) due to the distribution of critical temperature becoming increasingly skewed towards smaller values.
}\label{F_tcmat}
\end{centering}
\end{figure}

Here we compute $T_c^{\text{mat}}$ assuming that the superconductor-insulator transition is driven by a percolation transition. 
We define the percolation threshold as the temperature for which the area fraction $\phi$ of the sample which is above its local critical temperature $T_c({\bf r_0})$ is $\phi\to \phi_c=0.676 $\cite{Quintanilla2007}. The model is that of a two dimensional surface where circular superconducting regions form at random positions. The percolation transition occurs when there is sufficient superconducting area 
 that there exists a superconducting region which completely traverses the surface. The critical temperature of the material $T_c^{\text{mat}}$ is thus defined as,
\begin{equation}\label{tcper}
\int_0^{T_c^{\text{mat}}}\mathcal{P}(T_c({\bf r})) d T_c({\bf r})=1-\phi_c.
\end{equation}
The decrease in the mean value of $\mathcal{P}\left(\frac{T_c({\bf r})}{T_{c_\gamma}}\right)$ with increasing disorder suppresses the large exponential enhancement of $\Delta_\gamma$. The enhancement of the material bulk critical temperature is always much smaller than that of $\Delta_\gamma$, see figure \ref{F_tcmat}. It is only substantially higher than for non-disordered samples in the limit of very small electron-phonon coupling constant that might still describe materials like aluminium. In all other cases a very modest or no enhancement at all is observed.

\section{Estimation of the reduction of the critical temperature due to phase fluctuations}
\begin{figure}[h!]
\begin{centering}
\includegraphics[width=\wid\textwidth]{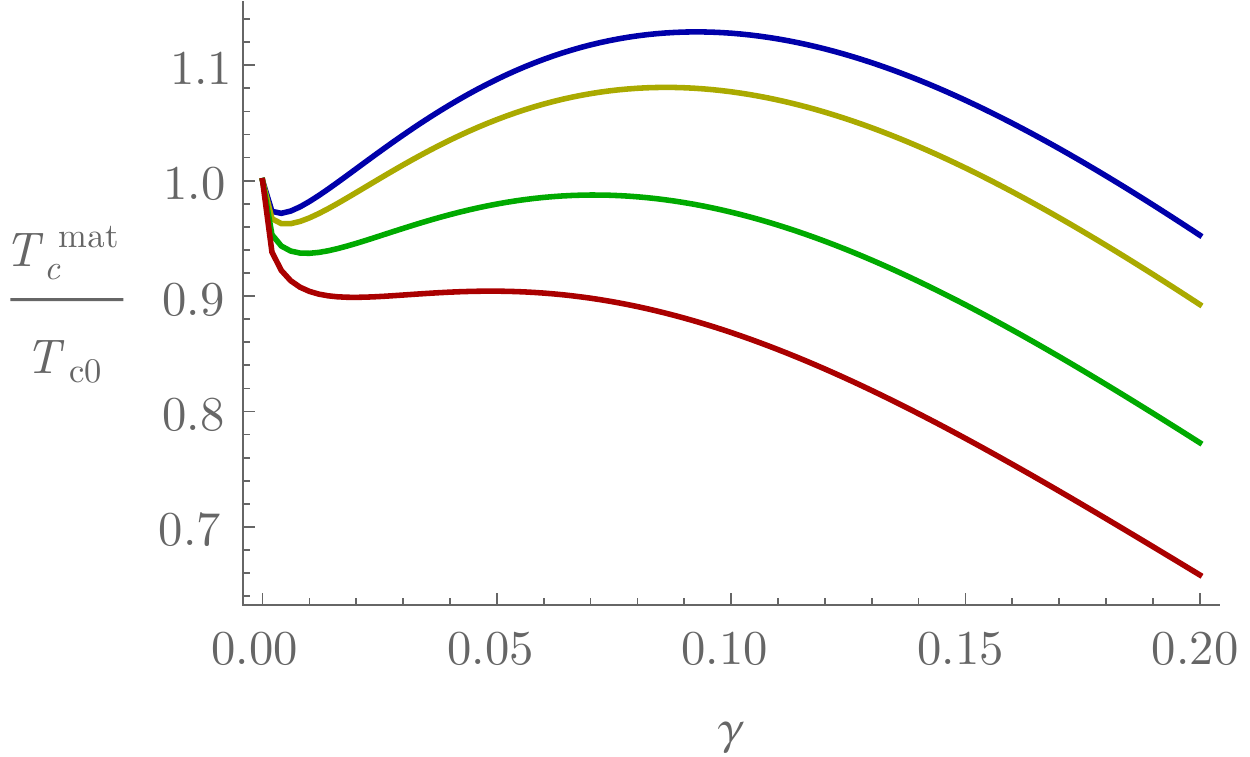}
\includegraphics[width=\wid\textwidth]{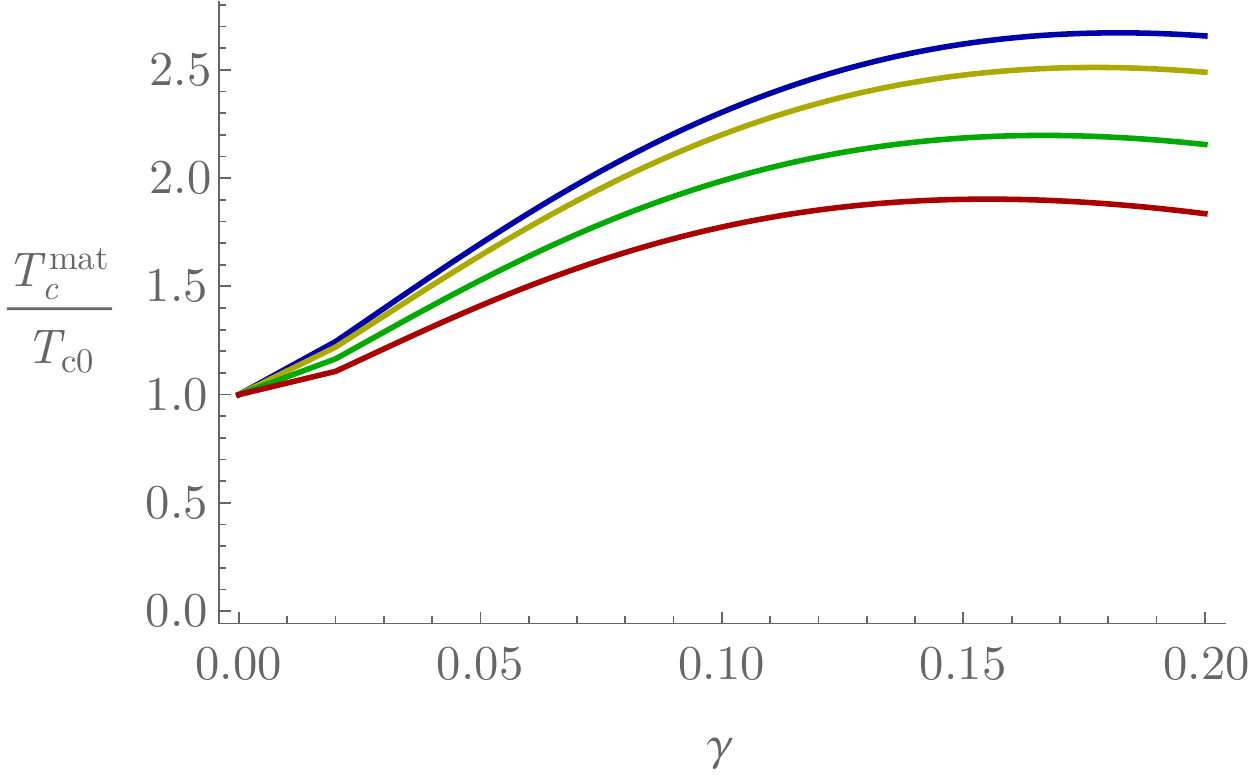}
\caption{The global critical temperature, $T_c^{\text{mat}}$, from Eqs. (\ref{PD}) and (\ref{tcper}), 
 in units of the clean critical temperature, as a function of the multifractal exponent $\gamma$ for $E_0/\epsilon_D=100$, $\lambda=0.4$ ({\it Upper plot}) and $\lambda=0.25$ ({\it Lower plot}) at the percolation threshold $\phi_c=0.676$ (Blue), and above it, $0.7$(Yellow), $0.75$(Green), $0.8$(Red). An area, $\phi^{Q}_c$, greater than the percolation threshold $\phi_c$, crudely mimics the effect of phase fluctuations that can break phase-coherence even above percolation threshold. The behaviour of $T_c^{\text{mat}}$ is strongly dependent on the choice of the critical area $\phi^{Q}_c$. We observe that the critical temperature decreases as $\phi^{Q}_c$ increases, which for $\lambda = 0.4$ rapidly suppresses any enhancement of the critical temperature with respect to the clean limit. By contrast for $\lambda = 0.25$ a substantial enhancement still occurs even for comparatively large values of $\phi^{Q}_c$. However this is still an upper bound of the enhancement that can be observed experimentally as we do not take into account the suppression of the order parameter amplitude induced by Coulomb interactions and other processes. Therefore we expect small or no enhancement except, possibly, for materials such as aluminium that are good metals and have very weak electron-phonon coupling.}\label{F_tcmat_phi}
\end{centering}
\end{figure}

 An obvious shortcoming of our model is the omission of Coulomb interactions and other sources of phase fluctuations that will reduce significantly the critical temperature of the sample as phase coherence can be lost even above the percolation threshold. 
Unfortunately a quantitative analytical estimation of these effects is in general quite hard.
Even the standard perturbative prediction \cite{Maekawa1982},
$
\frac{\delta T_c}{T_c} \sim \frac{\lambda_{\rm effec}}{g}{\rm ln}^2(\epsilon_D/T_c) 
$
for the decrease of $T_c$ leaves the final result in terms of the effective strength of the interaction $\lambda_{\rm effec}$ which is in general difficult to estimate especially in a disordered system. The recently developed formalism \cite{Mayoh2014a} to address arrays of superconducting nano-grains, that includes charging effects, could, at least qualitatively be adapted to this case. However it is difficult to estimate rigorously the capacitance in this context. Moreover we also neglect recombination processes of the order parameter and interactions with single quasi-particles. This is likely a good approximation 
for low temperatures but for higher temperatures closer to the critical one \cite{Kapitulnik1985} it is plausible that these processes will effectively broaden the Ginzburg region of the superconductor and further lower its critical temperature. Again for metallic superconductors it is difficult to make a fully quantitative estimation of the importance of these corrections. 
Despite these limitations it is clear that phase correlations persist only on an intricate network \cite{erez2013} above the percolation threshold for the amplitude of the order parameter \cite{Mayoh2014a}. At least qualitatively it seems therefore plausible that the true global critical temperature of the system $T_c^{\text{mat}}$, that includes the effect of phase fluctuations, can still be estimated by percolation techniques by increasing the percolation threshold. This method we apply here to estimate $T_c^{\text{mat}}$. For no phase fluctuations the global critical temperature is obtained by setting the fraction, $\phi$,  of the superconductor which is above the local critical temperature to the percolation threshold $\phi \approx \phi_c = 0.675$. Therefore the global critical temperature associated with larger values $\phi > \phi_c$ corresponds to situations where the superconducting fraction is sufficient to support a supercurrent but phase fluctuations prevent phase coherence. We expect the critical area, $\phi_c^Q$, in realistic situations to be higher than the percolation prediction
$\phi_c=0.676$. In figure \ref{F_tcmat_phi} we compare the global critical temperature for different values of $\phi_c^Q$, which roughly speaking model the effect of phase fluctuations, and the electron-phonon coupling $\lambda$. For sufficiently large $\lambda$ any enhancement at $\phi_c$ is rapidly suppressed with increasing disorder. By contrast for sufficiently small $\lambda$ the enhancement persists even for relatively large values of $\phi_c^Q$. We expect the trend of decreasing critical temperature to continue up to stronger disorder, which would agree with the experimental results \cite{Haviland1989,goldman1993}. It is important to stress that this method to mimic the effect of phase fluctuations does not take into account the fact that Coulomb interactions not only induce phase fluctuation but also decrease the superconducting gap and the local critical temperature. Therefore even the observed substantial enhancement for very weak coupling is only an upper bound of the one that could be observed experimentally. 


Clearly, a more refined model, beyond the scope of the paper, would be highly desirable to account quantitatively for the effect of phase fluctuations. However our results suggest that enhancement of the global critical temperature might be possible but only in very weakly coupled superconductors.
\section{Relevance to experiments}
Currently it is feasible to test some of the above theoretical predictions in disordered thin films. Scanning tunneling microscope techniques could be used to measure $\Delta(r_0)$ and $T_c(r_0)$ where the latter is experimentally defined as the temperature for which the gap in the differential conductance vanishes. Indeed the statistical distribution function of the gap, recently measured experimentally in strongly disordered Nb thin films \cite{Lemarie2013} close to the transition, seem qualitatively similar to the log-normal distribution that we have obtained analytically. However, for a quantitative comparison a higher resolution in the experimental results is necessary.
Our results could also be employed to measure the multifractal dimensions and the strength of disorder. For instance, according to Eq. (\ref{ratioexp}), the ratio between $\Delta(r_0)$ and $T_c(r_0)$ only depends on the multifractal exponent $\gamma$ and not on the coupling constant. Experimentally, it could be possible to average over $r_0$ to measure this ratio with better accuracy. 

Transport measurement like the resistivity could highlight the difference between the local critical temperature $T_c(r_0)$ and the global critical temperature defined as the highest temperature for which a supercurrent can flow. The latter should correspond with our prediction for the global critical temperature $T_c^{\text{mat}}$ resulting from the percolation analysis above. Indeed the sharpness of the transition as a function of the temperature could provide important clues on the role of phase fluctuations and percolation of the amplitude in the determination of the global critical temperature. 

Specific heat measurements would be a straightforward approach to studying the nature and properties of the phase transition. In particular the width and height of the peak would supply important information about the superconducting area fraction at the transition and about the distribution function $\mathcal{P}(T({\bf r}))$.

Finally we stress that one of the main results of the paper, that enhancement of $T_c^{\text{mat}}$ by disorder can only be observed in materials with a very weak electron-phonon coupling, is fully consistent with experimental results. It is well known \cite{Abeles1966,goldman1993} that the critical temperature of Al thin films start to increase as the thickness enters in the nano-scale region. By contrast in more strongly coupled superconductors like Pb no enhancement is observed \cite{Haviland1989,goldman1993} and the critical temperature decreases monotonically as the thickness decreases or the disorder strength increases. We note that as the thickness is decreased the material becomes quasi-two dimensional where multifractality is generic for sufficiently weak disorder. This is the case for metallic superconductors such as Al which are good conductors above the critical temperature.


\section{Conclusions}
We have studied the interplay between superconductiviy and disorder in a system characterized by weakly multifractal one-body eigenstates. This setting is especially appealing as multifractality enhances pairing correlations and induces strong spatial inhomogeneities in the superconducting order parameter but at the same time it is possible to obtain analytical results. First, we have computed exactly the superconducting gap at the Fermi energy, as a function of the multifractal dimensions, and the temperature at which it vanishes. We have found an enhancement of the gap with respect to the clean limit, but much smaller than in recent claims of the literature. Then, based on the calculation of the energy dependence of the order parameter, we have found that the order parameter is strongly inhomogeneous in space with a distribution function that follows a log-normal distribution. Interestingly the maximum of the distribution deviates strongly from the value of the gap at the Fermi energy as multifractality increases. This suggests that the global critical temperature of the superconductor, defined as the maximum temperature at which a supercurrent can flow, is much lower than the one found by considering the temperature at which the gap at the Fermi energy vanishes. In order to test this claim we employ percolation techniques to compute an upper-bound on the global critical temperature. Our formalism does not include directly phase fluctuations, induced by Coulomb interactions or other mechanisms, that further reduce the critical temperature. As a crude method to simulate these effects we have also computed the global critical temperature when the condition for percolation is slightly increased. The outcome of this analysis is that a substantial enhancement of the critical temperature might be possible only for very weak electron-phonon coupling. 
This could explain the well known experimental result \cite{Abeles1966,goldman1993} that in aluminium, a material with very weak electron-phonon coupling, the critical temperature is substantially enhanced with respect to the clean limit when the thickness of the sample is sufficiently small. In this limit the material is disordered and quasi-two dimensional so multifractality might play a role and our formalism is applicable. 
Our results are relevant to a number of physical situations including weakly disordered two dimensional systems of size much smaller than the localization length, bulk two dimensional disordered systems with spin-orbit interactions for which a metal insulator transition occurs in the weak disorder region and weakly disordered metals in $2+\epsilon$ dimensions with $\epsilon \ll 1$.
We hope this work will stimulate further experimental and theoretical research on superconductivity in disordered systems, especially the role of the Coulomb interaction in strongly inhomogeneous systems.

 \acknowledgments
 The authors would like to thank Lara Benfatto for useful discussions. AMG would like to thank Sangita Bose and Pratap Raychaudhuri for illuminating discussions. AMG was supported by EPSRC, grant No. EP/I004637/1, FCT, grant PTDC/FIS/111348/2009 and a Marie Curie International Reintegration Grant PIRG07-GA-2010-268172. JM acknowledges the support of an EPSRC PhD studentship.
 \vspace{1cm}
 
 \bibliography{library}

\begin{thebibliography}{54}%
\makeatletter
\providecommand \@ifxundefined [1]{%
 \@ifx{#1\undefined}
}%
\providecommand \@ifnum [1]{%
 \ifnum #1\expandafter \@firstoftwo
 \else \expandafter \@secondoftwo
 \fi
}%
\providecommand \@ifx [1]{%
 \ifx #1\expandafter \@firstoftwo
 \else \expandafter \@secondoftwo
 \fi
}%
\providecommand \natexlab [1]{#1}%
\providecommand \enquote  [1]{``#1''}%
\providecommand \bibnamefont  [1]{#1}%
\providecommand \bibfnamefont [1]{#1}%
\providecommand \citenamefont [1]{#1}%
\providecommand \href@noop [0]{\@secondoftwo}%
\providecommand \href [0]{\begingroup \@sanitize@url \@href}%
\providecommand \@href[1]{\@@startlink{#1}\@@href}%
\providecommand \@@href[1]{\endgroup#1\@@endlink}%
\providecommand \@sanitize@url [0]{\catcode `\\12\catcode `\$12\catcode
  `\&12\catcode `\#12\catcode `\^12\catcode `\_12\catcode `\%12\relax}%
\providecommand \@@startlink[1]{}%
\providecommand \@@endlink[0]{}%
\providecommand \url  [0]{\begingroup\@sanitize@url \@url }%
\providecommand \@url [1]{\endgroup\@href {#1}{\urlprefix }}%
\providecommand \urlprefix  [0]{URL }%
\providecommand \Eprint [0]{\href }%
\providecommand \doibase [0]{http://dx.doi.org/}%
\providecommand \selectlanguage [0]{\@gobble}%
\providecommand \bibinfo  [0]{\@secondoftwo}%
\providecommand \bibfield  [0]{\@secondoftwo}%
\providecommand \translation [1]{[#1]}%
\providecommand \BibitemOpen [0]{}%
\providecommand \bibitemStop [0]{}%
\providecommand \bibitemNoStop [0]{.\EOS\space}%
\providecommand \EOS [0]{\spacefactor3000\relax}%
\providecommand \BibitemShut  [1]{\csname bibitem#1\endcsname}%
\let\auto@bib@innerbib\@empty
\bibitem [{\citenamefont {Anderson}(1959)}]{Anderson1959}%
  \BibitemOpen
  \bibfield  {author} {\bibinfo {author} {\bibfnamefont {P.~W.}\ \bibnamefont
  {Anderson}},\ }\href
  {http://www.sciencedirect.com/science/article/pii/0022369759900368}
  {\bibfield  {journal} {\bibinfo  {journal} {J. Phys. Chem. Solids}\ }\textbf
  {\bibinfo {volume} {11}},\ \bibinfo {pages} {26} (\bibinfo {year}
  {1959})}\BibitemShut {NoStop}%
\bibitem [{\citenamefont {Abrikosov}\ and\ \citenamefont
  {Gorkov}(1961)}]{Abrikosov1961}%
  \BibitemOpen
  \bibfield  {author} {\bibinfo {author} {\bibfnamefont {A.~A.}\ \bibnamefont
  {Abrikosov}}\ and\ \bibinfo {author} {\bibfnamefont {L.~P.}\ \bibnamefont
  {Gorkov}},\ }\href@noop {} {\bibfield  {journal} {\bibinfo  {journal} {J.
  Exp. Theor. Phys.}\ }\textbf {\bibinfo {volume} {12}},\ \bibinfo {pages}
  {1243} (\bibinfo {year} {1961})}\BibitemShut {NoStop}%
\bibitem [{\citenamefont {Kim}\ and\ \citenamefont
  {Overhauser}(1993)}]{Kim1993}%
  \BibitemOpen
  \bibfield  {author} {\bibinfo {author} {\bibfnamefont {Y.-J.}\ \bibnamefont
  {Kim}}\ and\ \bibinfo {author} {\bibfnamefont {A.}~\bibnamefont
  {Overhauser}},\ }\href {\doibase 10.1103/PhysRevB.47.8025} {\bibfield
  {journal} {\bibinfo  {journal} {Phys. Rev. B}\ }\textbf {\bibinfo {volume}
  {47}},\ \bibinfo {pages} {8025} (\bibinfo {year} {1993})}\BibitemShut
  {NoStop}%
\bibitem [{\citenamefont {Abrikosov}\ and\ \citenamefont
  {Gor’kov}(1994)}]{Abrikosov1994}%
  \BibitemOpen
  \bibfield  {author} {\bibinfo {author} {\bibfnamefont {A.}~\bibnamefont
  {Abrikosov}}\ and\ \bibinfo {author} {\bibfnamefont {L.}~\bibnamefont
  {Gor’kov}},\ }\href {\doibase 10.1103/PhysRevB.49.12337} {\bibfield
  {journal} {\bibinfo  {journal} {Phys. Rev. B}\ }\textbf {\bibinfo {volume}
  {49}},\ \bibinfo {pages} {12337} (\bibinfo {year} {1994})}\BibitemShut
  {NoStop}%
\bibitem [{\citenamefont {de~Gennes}(1964)}]{DeGennes1964}%
  \BibitemOpen
  \bibfield  {author} {\bibinfo {author} {\bibfnamefont {P.}~\bibnamefont
  {de~Gennes}},\ }\href {\doibase 10.1103/RevModPhys.36.225} {\bibfield
  {journal} {\bibinfo  {journal} {Rev. Mod. Phys.}\ }\textbf {\bibinfo {volume}
  {36}},\ \bibinfo {pages} {225} (\bibinfo {year} {1964})}\BibitemShut
  {NoStop}%
\bibitem [{\citenamefont {Haviland}\ \emph {et~al.}(1989)\citenamefont
  {Haviland}, \citenamefont {Liu},\ and\ \citenamefont
  {Goldman}}]{Haviland1989}%
  \BibitemOpen
  \bibfield  {author} {\bibinfo {author} {\bibfnamefont {D.~B.}\ \bibnamefont
  {Haviland}}, \bibinfo {author} {\bibfnamefont {Y.}~\bibnamefont {Liu}}, \
  and\ \bibinfo {author} {\bibfnamefont {A.~M.}\ \bibnamefont {Goldman}},\
  }\href {\doibase 10.1103/PhysRevLett.62.2180} {\bibfield  {journal} {\bibinfo
   {journal} {Phys. Rev. Lett.}\ }\textbf {\bibinfo {volume} {62}},\ \bibinfo
  {pages} {2180} (\bibinfo {year} {1989})}\BibitemShut {NoStop}%
\bibitem [{\citenamefont {Liu}\ \emph {et~al.}(1993)\citenamefont {Liu},
  \citenamefont {Haviland}, \citenamefont {Nease},\ and\ \citenamefont
  {Goldman}}]{goldman1993}%
  \BibitemOpen
  \bibfield  {author} {\bibinfo {author} {\bibfnamefont {Y.}~\bibnamefont
  {Liu}}, \bibinfo {author} {\bibfnamefont {D.~B.}\ \bibnamefont {Haviland}},
  \bibinfo {author} {\bibfnamefont {B.}~\bibnamefont {Nease}}, \ and\ \bibinfo
  {author} {\bibfnamefont {A.~M.}\ \bibnamefont {Goldman}},\ }\href {\doibase
  10.1103/PhysRevB.47.5931} {\bibfield  {journal} {\bibinfo  {journal} {Phys.
  Rev. B}\ }\textbf {\bibinfo {volume} {47}},\ \bibinfo {pages} {5931}
  (\bibinfo {year} {1993})}\BibitemShut {NoStop}%
\bibitem [{\citenamefont {Nishida}\ \emph {et~al.}(1982)\citenamefont
  {Nishida}, \citenamefont {Yamaguchi}, \citenamefont {Furubayashi},
  \citenamefont {Morigaki}, \citenamefont {Ishimoto},\ and\ \citenamefont
  {Ono}}]{Nishida1982}%
  \BibitemOpen
  \bibfield  {author} {\bibinfo {author} {\bibfnamefont {N.}~\bibnamefont
  {Nishida}}, \bibinfo {author} {\bibfnamefont {M.}~\bibnamefont {Yamaguchi}},
  \bibinfo {author} {\bibfnamefont {T.}~\bibnamefont {Furubayashi}}, \bibinfo
  {author} {\bibfnamefont {K.}~\bibnamefont {Morigaki}}, \bibinfo {author}
  {\bibfnamefont {H.}~\bibnamefont {Ishimoto}}, \ and\ \bibinfo {author}
  {\bibfnamefont {K.}~\bibnamefont {Ono}},\ }\href {\doibase
  10.1016/0038-1098(82)90455-0} {\bibfield  {journal} {\bibinfo  {journal}
  {Solid State Commun.}\ }\textbf {\bibinfo {volume} {44}},\ \bibinfo {pages}
  {305} (\bibinfo {year} {1982})}\BibitemShut {NoStop}%
\bibitem [{\citenamefont {Furubayashi}\ \emph {et~al.}(1985)\citenamefont
  {Furubayashi}, \citenamefont {Nishida}, \citenamefont {Yamaguchi},
  \citenamefont {Morigaki},\ and\ \citenamefont {Ishimoto}}]{Furubayashi1985}%
  \BibitemOpen
  \bibfield  {author} {\bibinfo {author} {\bibfnamefont {T.}~\bibnamefont
  {Furubayashi}}, \bibinfo {author} {\bibfnamefont {N.}~\bibnamefont
  {Nishida}}, \bibinfo {author} {\bibfnamefont {M.}~\bibnamefont {Yamaguchi}},
  \bibinfo {author} {\bibfnamefont {K.}~\bibnamefont {Morigaki}}, \ and\
  \bibinfo {author} {\bibfnamefont {H.}~\bibnamefont {Ishimoto}},\ }\href
  {\doibase 10.1016/0038-1098(85)90324-2} {\bibfield  {journal} {\bibinfo
  {journal} {Solid State Commun.}\ }\textbf {\bibinfo {volume} {55}},\ \bibinfo
  {pages} {513} (\bibinfo {year} {1985})}\BibitemShut {NoStop}%
\bibitem [{\citenamefont {Alekseevskii}\ \emph {et~al.}(1983)\citenamefont
  {Alekseevskii}, \citenamefont {Mitin}, \citenamefont {Samosyuk},\ and\
  \citenamefont {Firsov}}]{Alekseevskii1983}%
  \BibitemOpen
  \bibfield  {author} {\bibinfo {author} {\bibfnamefont {N.}~\bibnamefont
  {Alekseevskii}}, \bibinfo {author} {\bibfnamefont {A.}~\bibnamefont {Mitin}},
  \bibinfo {author} {\bibfnamefont {V.}~\bibnamefont {Samosyuk}}, \ and\
  \bibinfo {author} {\bibfnamefont {V.}~\bibnamefont {Firsov}},\ }\href@noop {}
  {\bibfield  {journal} {\bibinfo  {journal} {J. Exp. Theor. Phys.}\ }\textbf
  {\bibinfo {volume} {58}},\ \bibinfo {pages} {635} (\bibinfo {year}
  {1983})}\BibitemShut {NoStop}%
\bibitem [{\citenamefont {Bishop}\ \emph {et~al.}(1985)\citenamefont {Bishop},
  \citenamefont {Spencer},\ and\ \citenamefont {Dynes}}]{Bishop1985}%
  \BibitemOpen
  \bibfield  {author} {\bibinfo {author} {\bibfnamefont {D.}~\bibnamefont
  {Bishop}}, \bibinfo {author} {\bibfnamefont {E.}~\bibnamefont {Spencer}}, \
  and\ \bibinfo {author} {\bibfnamefont {R.}~\bibnamefont {Dynes}},\ }\href
  {\doibase 10.1016/0038-1101(85)90212-6} {\bibfield  {journal} {\bibinfo
  {journal} {Solid. State. Electron.}\ }\textbf {\bibinfo {volume} {28}},\
  \bibinfo {pages} {73} (\bibinfo {year} {1985})}\BibitemShut {NoStop}%
\bibitem [{\citenamefont {Graybeal}(1985)}]{Graybeal1985}%
  \BibitemOpen
  \bibfield  {author} {\bibinfo {author} {\bibfnamefont {J.}~\bibnamefont
  {Graybeal}},\ }\href {\doibase 10.1016/0378-4363(85)90448-6} {\bibfield
  {journal} {\bibinfo  {journal} {Phys. B+C}\ }\textbf {\bibinfo {volume}
  {135}},\ \bibinfo {pages} {113} (\bibinfo {year} {1985})}\BibitemShut
  {NoStop}%
\bibitem [{\citenamefont {Driessen}\ \emph {et~al.}(2012)\citenamefont
  {Driessen}, \citenamefont {Coumou}, \citenamefont {Tromp}, \citenamefont
  {de~Visser},\ and\ \citenamefont {Klapwijk}}]{Driessen2012}%
  \BibitemOpen
  \bibfield  {author} {\bibinfo {author} {\bibfnamefont {E.~F.~C.}\
  \bibnamefont {Driessen}}, \bibinfo {author} {\bibfnamefont {P.~C. J.~J.}\
  \bibnamefont {Coumou}}, \bibinfo {author} {\bibfnamefont {R.~R.}\
  \bibnamefont {Tromp}}, \bibinfo {author} {\bibfnamefont {P.~J.}\ \bibnamefont
  {de~Visser}}, \ and\ \bibinfo {author} {\bibfnamefont {T.~M.}\ \bibnamefont
  {Klapwijk}},\ }\href {\doibase 10.1103/PhysRevLett.109.107003} {\bibfield
  {journal} {\bibinfo  {journal} {Phys. Rev. Lett.}\ }\textbf {\bibinfo
  {volume} {109}},\ \bibinfo {pages} {107003} (\bibinfo {year}
  {2012})}\BibitemShut {NoStop}%
\bibitem [{\citenamefont {Tashiro}\ \emph {et~al.}(2008)\citenamefont
  {Tashiro}, \citenamefont {Graybeal}, \citenamefont {Tanner}, \citenamefont
  {Nicol}, \citenamefont {Carbotte},\ and\ \citenamefont {Carr}}]{Tashiro2008}%
  \BibitemOpen
  \bibfield  {author} {\bibinfo {author} {\bibfnamefont {H.}~\bibnamefont
  {Tashiro}}, \bibinfo {author} {\bibfnamefont {J.}~\bibnamefont {Graybeal}},
  \bibinfo {author} {\bibfnamefont {D.}~\bibnamefont {Tanner}}, \bibinfo
  {author} {\bibfnamefont {E.}~\bibnamefont {Nicol}}, \bibinfo {author}
  {\bibfnamefont {J.}~\bibnamefont {Carbotte}}, \ and\ \bibinfo {author}
  {\bibfnamefont {G.}~\bibnamefont {Carr}},\ }\href {\doibase
  10.1103/PhysRevB.78.014509} {\bibfield  {journal} {\bibinfo  {journal} {Phys.
  Rev. B}\ }\textbf {\bibinfo {volume} {78}},\ \bibinfo {pages} {014509}
  (\bibinfo {year} {2008})}\BibitemShut {NoStop}%
\bibitem [{\citenamefont {Maekawa}\ and\ \citenamefont
  {Fukuyama}(1982)}]{Maekawa1982}%
  \BibitemOpen
  \bibfield  {author} {\bibinfo {author} {\bibfnamefont {S.}~\bibnamefont
  {Maekawa}}\ and\ \bibinfo {author} {\bibfnamefont {H.}~\bibnamefont
  {Fukuyama}},\ }\href@noop {} {\bibfield  {journal} {\bibinfo  {journal} {J.
  Phys. Soc. Japan}\ }\textbf {\bibinfo {volume} {51}},\ \bibinfo {pages}
  {1380} (\bibinfo {year} {1982})}\BibitemShut {NoStop}%
\bibitem [{\citenamefont {Maekawa}\ \emph {et~al.}(1984)\citenamefont
  {Maekawa}, \citenamefont {Ebisawa},\ and\ \citenamefont
  {Fukuyama}}]{Maekawa1984}%
  \BibitemOpen
  \bibfield  {author} {\bibinfo {author} {\bibfnamefont {S.}~\bibnamefont
  {Maekawa}}, \bibinfo {author} {\bibfnamefont {H.}~\bibnamefont {Ebisawa}}, \
  and\ \bibinfo {author} {\bibfnamefont {H.}~\bibnamefont {Fukuyama}},\
  }\href@noop {} {\bibfield  {journal} {\bibinfo  {journal} {J. Phys. Soc.
  Japan}\ }\textbf {\bibinfo {volume} {53}},\ \bibinfo {pages} {2681} (\bibinfo
  {year} {1984})}\BibitemShut {NoStop}%
\bibitem [{\citenamefont {Ghosal}\ \emph {et~al.}(2001)\citenamefont {Ghosal},
  \citenamefont {Randeria},\ and\ \citenamefont {Trivedi}}]{Ghosal2001}%
  \BibitemOpen
  \bibfield  {author} {\bibinfo {author} {\bibfnamefont {A.}~\bibnamefont
  {Ghosal}}, \bibinfo {author} {\bibfnamefont {M.}~\bibnamefont {Randeria}}, \
  and\ \bibinfo {author} {\bibfnamefont {N.}~\bibnamefont {Trivedi}},\ }\href
  {\doibase 10.1103/PhysRevB.65.014501} {\bibfield  {journal} {\bibinfo
  {journal} {Phys. Rev. B}\ }\textbf {\bibinfo {volume} {65}},\ \bibinfo
  {pages} {014501} (\bibinfo {year} {2001})}\BibitemShut {NoStop}%
\bibitem [{\citenamefont {Bouadim}\ \emph {et~al.}(2011)\citenamefont
  {Bouadim}, \citenamefont {Loh}, \citenamefont {Randeria},\ and\ \citenamefont
  {Trivedi}}]{Bouadim2011}%
  \BibitemOpen
  \bibfield  {author} {\bibinfo {author} {\bibfnamefont {K.}~\bibnamefont
  {Bouadim}}, \bibinfo {author} {\bibfnamefont {Y.~L.}\ \bibnamefont {Loh}},
  \bibinfo {author} {\bibfnamefont {M.}~\bibnamefont {Randeria}}, \ and\
  \bibinfo {author} {\bibfnamefont {N.}~\bibnamefont {Trivedi}},\ }\href@noop
  {} {\bibfield  {journal} {\bibinfo  {journal} {Nat. Phys.}\ }\textbf
  {\bibinfo {volume} {7}},\ \bibinfo {pages} {884} (\bibinfo {year}
  {2011})}\BibitemShut {NoStop}%
\bibitem [{\citenamefont {Mondal}\ \emph {et~al.}(2011)\citenamefont {Mondal},
  \citenamefont {Kamlapure}, \citenamefont {Chand}, \citenamefont {Saraswat},
  \citenamefont {Kumar}, \citenamefont {Jesudasan}, \citenamefont {Benfatto},
  \citenamefont {Tripathi},\ and\ \citenamefont {Raychaudhuri}}]{Mondal2011}%
  \BibitemOpen
  \bibfield  {author} {\bibinfo {author} {\bibfnamefont {M.}~\bibnamefont
  {Mondal}}, \bibinfo {author} {\bibfnamefont {A.}~\bibnamefont {Kamlapure}},
  \bibinfo {author} {\bibfnamefont {M.}~\bibnamefont {Chand}}, \bibinfo
  {author} {\bibfnamefont {G.}~\bibnamefont {Saraswat}}, \bibinfo {author}
  {\bibfnamefont {S.}~\bibnamefont {Kumar}}, \bibinfo {author} {\bibfnamefont
  {J.}~\bibnamefont {Jesudasan}}, \bibinfo {author} {\bibfnamefont
  {L.}~\bibnamefont {Benfatto}}, \bibinfo {author} {\bibfnamefont
  {V.}~\bibnamefont {Tripathi}}, \ and\ \bibinfo {author} {\bibfnamefont
  {P.}~\bibnamefont {Raychaudhuri}},\ }\href@noop {} {\bibfield  {journal}
  {\bibinfo  {journal} {Phys. Rev. Lett.}\ }\textbf {\bibinfo {volume} {106}},\
  \bibinfo {pages} {47001} (\bibinfo {year} {2011})}\BibitemShut {NoStop}%
\bibitem [{\citenamefont {Trivedi}\ \emph {et~al.}(2012)\citenamefont
  {Trivedi}, \citenamefont {Loh}, \citenamefont {Bouadim},\ and\ \citenamefont
  {Randeria}}]{Trivedi2012}%
  \BibitemOpen
  \bibfield  {author} {\bibinfo {author} {\bibfnamefont {N.}~\bibnamefont
  {Trivedi}}, \bibinfo {author} {\bibfnamefont {Y.~L.}\ \bibnamefont {Loh}},
  \bibinfo {author} {\bibfnamefont {K.}~\bibnamefont {Bouadim}}, \ and\
  \bibinfo {author} {\bibfnamefont {M.}~\bibnamefont {Randeria}},\ }\href
  {http://stacks.iop.org/1742-6596/376/i=1/a=012001} {\bibfield  {journal}
  {\bibinfo  {journal} {J. Phys. Conf. Ser.}\ }\textbf {\bibinfo {volume}
  {376}},\ \bibinfo {pages} {12001} (\bibinfo {year} {2012})}\BibitemShut
  {NoStop}%
\bibitem [{\citenamefont {Sherman}\ \emph {et~al.}(2014)\citenamefont
  {Sherman}, \citenamefont {Gorshunov}, \citenamefont {Poran}, \citenamefont
  {Trivedi}, \citenamefont {Farber}, \citenamefont {Dressel},\ and\
  \citenamefont {Frydman}}]{Sherman2014}%
  \BibitemOpen
  \bibfield  {author} {\bibinfo {author} {\bibfnamefont {D.}~\bibnamefont
  {Sherman}}, \bibinfo {author} {\bibfnamefont {B.}~\bibnamefont {Gorshunov}},
  \bibinfo {author} {\bibfnamefont {S.}~\bibnamefont {Poran}}, \bibinfo
  {author} {\bibfnamefont {N.}~\bibnamefont {Trivedi}}, \bibinfo {author}
  {\bibfnamefont {E.}~\bibnamefont {Farber}}, \bibinfo {author} {\bibfnamefont
  {M.}~\bibnamefont {Dressel}}, \ and\ \bibinfo {author} {\bibfnamefont
  {A.}~\bibnamefont {Frydman}},\ }\href {\doibase 10.1103/PhysRevB.89.035149}
  {\bibfield  {journal} {\bibinfo  {journal} {Phys. Rev. B}\ }\textbf {\bibinfo
  {volume} {89}},\ \bibinfo {pages} {35149} (\bibinfo {year}
  {2014})}\BibitemShut {NoStop}%
\bibitem [{\citenamefont {Erez}\ and\ \citenamefont {Meir}(2013)}]{erez2013}%
  \BibitemOpen
  \bibfield  {author} {\bibinfo {author} {\bibfnamefont {A.}~\bibnamefont
  {Erez}}\ and\ \bibinfo {author} {\bibfnamefont {Y.}~\bibnamefont {Meir}},\
  }\href {\doibase 10.1103/PhysRevLett.111.187002} {\bibfield  {journal}
  {\bibinfo  {journal} {Phys. Rev. Lett.}\ }\textbf {\bibinfo {volume} {111}},\
  \bibinfo {pages} {187002} (\bibinfo {year} {2013})}\BibitemShut {NoStop}%
\bibitem [{\citenamefont {Lemari\'{e}}\ \emph {et~al.}(2013)\citenamefont
  {Lemari\'{e}}, \citenamefont {Kamlapure}, \citenamefont {Bucheli},
  \citenamefont {Benfatto}, \citenamefont {Lorenzana}, \citenamefont {Seibold},
  \citenamefont {Ganguli}, \citenamefont {Raychaudhuri},\ and\ \citenamefont
  {Castellani}}]{Lemarie2013}%
  \BibitemOpen
  \bibfield  {author} {\bibinfo {author} {\bibfnamefont {G.}~\bibnamefont
  {Lemari\'{e}}}, \bibinfo {author} {\bibfnamefont {A.}~\bibnamefont
  {Kamlapure}}, \bibinfo {author} {\bibfnamefont {D.}~\bibnamefont {Bucheli}},
  \bibinfo {author} {\bibfnamefont {L.}~\bibnamefont {Benfatto}}, \bibinfo
  {author} {\bibfnamefont {J.}~\bibnamefont {Lorenzana}}, \bibinfo {author}
  {\bibfnamefont {G.}~\bibnamefont {Seibold}}, \bibinfo {author} {\bibfnamefont
  {S.~C.}\ \bibnamefont {Ganguli}}, \bibinfo {author} {\bibfnamefont
  {P.}~\bibnamefont {Raychaudhuri}}, \ and\ \bibinfo {author} {\bibfnamefont
  {C.}~\bibnamefont {Castellani}},\ }\href {\doibase
  10.1103/PhysRevB.87.184509} {\bibfield  {journal} {\bibinfo  {journal} {Phys.
  Rev. B}\ }\textbf {\bibinfo {volume} {87}},\ \bibinfo {pages} {184509}
  (\bibinfo {year} {2013})}\BibitemShut {NoStop}%
\bibitem [{\citenamefont {Brun}\ \emph {et~al.}(2014)\citenamefont {Brun},
  \citenamefont {Cren}, \citenamefont {Cherkez}, \citenamefont {Debontridder},
  \citenamefont {Pons}, \citenamefont {Fokin}, \citenamefont {Tringides},
  \citenamefont {Bozhko}, \citenamefont {Ioffe}, \citenamefont {Altshuler}
  \emph {et~al.}}]{brun2014}%
  \BibitemOpen
  \bibfield  {author} {\bibinfo {author} {\bibfnamefont {C.}~\bibnamefont
  {Brun}}, \bibinfo {author} {\bibfnamefont {T.}~\bibnamefont {Cren}}, \bibinfo
  {author} {\bibfnamefont {V.}~\bibnamefont {Cherkez}}, \bibinfo {author}
  {\bibfnamefont {F.}~\bibnamefont {Debontridder}}, \bibinfo {author}
  {\bibfnamefont {S.}~\bibnamefont {Pons}}, \bibinfo {author} {\bibfnamefont
  {D.}~\bibnamefont {Fokin}}, \bibinfo {author} {\bibfnamefont
  {M.}~\bibnamefont {Tringides}}, \bibinfo {author} {\bibfnamefont
  {S.}~\bibnamefont {Bozhko}}, \bibinfo {author} {\bibfnamefont
  {L.}~\bibnamefont {Ioffe}}, \bibinfo {author} {\bibfnamefont
  {B.}~\bibnamefont {Altshuler}},  \emph {et~al.},\ }\href {\doibase
  10.1038/nphys2937} {\bibfield  {journal} {\bibinfo  {journal} {Nature
  Physics}\ }\textbf {\bibinfo {volume} {10}},\ \bibinfo {pages} {444}
  (\bibinfo {year} {2014})}\BibitemShut {NoStop}%
\bibitem [{\citenamefont {Noat}\ \emph {et~al.}(2013)\citenamefont {Noat},
  \citenamefont {Cherkez}, \citenamefont {Brun}, \citenamefont {Cren},
  \citenamefont {Carbillet}, \citenamefont {Debontridder}, \citenamefont
  {Ilin}, \citenamefont {Siegel}, \citenamefont {Semenov}, \citenamefont
  {H\"ubers},\ and\ \citenamefont {Roditchev}}]{noat2013}%
  \BibitemOpen
  \bibfield  {author} {\bibinfo {author} {\bibfnamefont {Y.}~\bibnamefont
  {Noat}}, \bibinfo {author} {\bibfnamefont {V.}~\bibnamefont {Cherkez}},
  \bibinfo {author} {\bibfnamefont {C.}~\bibnamefont {Brun}}, \bibinfo {author}
  {\bibfnamefont {T.}~\bibnamefont {Cren}}, \bibinfo {author} {\bibfnamefont
  {C.}~\bibnamefont {Carbillet}}, \bibinfo {author} {\bibfnamefont
  {F.}~\bibnamefont {Debontridder}}, \bibinfo {author} {\bibfnamefont
  {K.}~\bibnamefont {Ilin}}, \bibinfo {author} {\bibfnamefont {M.}~\bibnamefont
  {Siegel}}, \bibinfo {author} {\bibfnamefont {A.}~\bibnamefont {Semenov}},
  \bibinfo {author} {\bibfnamefont {H.-W.}\ \bibnamefont {H\"ubers}}, \ and\
  \bibinfo {author} {\bibfnamefont {D.}~\bibnamefont {Roditchev}},\ }\href
  {\doibase 10.1103/PhysRevB.88.014503} {\bibfield  {journal} {\bibinfo
  {journal} {Phys. Rev. B}\ }\textbf {\bibinfo {volume} {88}},\ \bibinfo
  {pages} {014503} (\bibinfo {year} {2013})}\BibitemShut {NoStop}%
\bibitem [{\citenamefont {Ioffe}\ and\ \citenamefont
  {M\'{e}zard}(2010)}]{Ioffe2010}%
  \BibitemOpen
  \bibfield  {author} {\bibinfo {author} {\bibfnamefont {L.~B.}\ \bibnamefont
  {Ioffe}}\ and\ \bibinfo {author} {\bibfnamefont {M.}~\bibnamefont
  {M\'{e}zard}},\ }\href {\doibase 10.1103/PhysRevLett.105.037001} {\bibfield
  {journal} {\bibinfo  {journal} {Phys. Rev. Lett.}\ }\textbf {\bibinfo
  {volume} {105}},\ \bibinfo {pages} {37001} (\bibinfo {year}
  {2010})}\BibitemShut {NoStop}%
\bibitem [{\citenamefont {Seibold}\ \emph {et~al.}(2012)\citenamefont
  {Seibold}, \citenamefont {Benfatto}, \citenamefont {Castellani},\ and\
  \citenamefont {Lorenzana}}]{Seibold2012}%
  \BibitemOpen
  \bibfield  {author} {\bibinfo {author} {\bibfnamefont {G.}~\bibnamefont
  {Seibold}}, \bibinfo {author} {\bibfnamefont {L.}~\bibnamefont {Benfatto}},
  \bibinfo {author} {\bibfnamefont {C.}~\bibnamefont {Castellani}}, \ and\
  \bibinfo {author} {\bibfnamefont {J.}~\bibnamefont {Lorenzana}},\ }\href
  {\doibase 10.1103/PhysRevLett.108.207004} {\bibfield  {journal} {\bibinfo
  {journal} {Phys. Rev. Lett.}\ }\textbf {\bibinfo {volume} {108}},\ \bibinfo
  {pages} {207004} (\bibinfo {year} {2012})}\BibitemShut {NoStop}%
\bibitem [{\citenamefont {Mondal}\ \emph {et~al.}(2013)\citenamefont {Mondal},
  \citenamefont {Kamlapure}, \citenamefont {Ganguli}, \citenamefont
  {Jesudasan}, \citenamefont {Bagwe}, \citenamefont {Benfatto},\ and\
  \citenamefont {Raychaudhuri}}]{Mondal2013}%
  \BibitemOpen
  \bibfield  {author} {\bibinfo {author} {\bibfnamefont {M.}~\bibnamefont
  {Mondal}}, \bibinfo {author} {\bibfnamefont {A.}~\bibnamefont {Kamlapure}},
  \bibinfo {author} {\bibfnamefont {S.~C.}\ \bibnamefont {Ganguli}}, \bibinfo
  {author} {\bibfnamefont {J.}~\bibnamefont {Jesudasan}}, \bibinfo {author}
  {\bibfnamefont {V.}~\bibnamefont {Bagwe}}, \bibinfo {author} {\bibfnamefont
  {L.}~\bibnamefont {Benfatto}}, \ and\ \bibinfo {author} {\bibfnamefont
  {P.}~\bibnamefont {Raychaudhuri}},\ }\href@noop {} {\bibfield  {journal}
  {\bibinfo  {journal} {Sci. Rep.}\ }\textbf {\bibinfo {volume} {3}} (\bibinfo
  {year} {2013})}\BibitemShut {NoStop}%
\bibitem [{\citenamefont {Sac\'{e}p\'{e}}\ \emph {et~al.}(2011)\citenamefont
  {Sac\'{e}p\'{e}}, \citenamefont {Dubouchet}, \citenamefont {Chapelier},
  \citenamefont {Sanquer}, \citenamefont {Ovadia}, \citenamefont {Shahar},
  \citenamefont {Feigel'man},\ and\ \citenamefont {Ioffe}}]{Sacepe2011}%
  \BibitemOpen
  \bibfield  {author} {\bibinfo {author} {\bibfnamefont {B.}~\bibnamefont
  {Sac\'{e}p\'{e}}}, \bibinfo {author} {\bibfnamefont {T.}~\bibnamefont
  {Dubouchet}}, \bibinfo {author} {\bibfnamefont {C.}~\bibnamefont
  {Chapelier}}, \bibinfo {author} {\bibfnamefont {M.}~\bibnamefont {Sanquer}},
  \bibinfo {author} {\bibfnamefont {M.}~\bibnamefont {Ovadia}}, \bibinfo
  {author} {\bibfnamefont {D.}~\bibnamefont {Shahar}}, \bibinfo {author}
  {\bibfnamefont {M.}~\bibnamefont {Feigel'man}}, \ and\ \bibinfo {author}
  {\bibfnamefont {L.}~\bibnamefont {Ioffe}},\ }\href {\doibase
  10.1038/nphys1892} {\bibfield  {journal} {\bibinfo  {journal} {Nat. Phys.}\
  }\textbf {\bibinfo {volume} {7}},\ \bibinfo {pages} {239} (\bibinfo {year}
  {2011})}\BibitemShut {NoStop}%
\bibitem [{\citenamefont {Suslov}(2013)}]{suslov2013}%
  \BibitemOpen
  \bibfield  {author} {\bibinfo {author} {\bibfnamefont {I.}~\bibnamefont
  {Suslov}},\ }\href@noop {} {\bibfield  {journal} {\bibinfo  {journal}
  {Journal of Experimental and Theoretical Physics}\ }\textbf {\bibinfo
  {volume} {117}},\ \bibinfo {pages} {1042} (\bibinfo {year}
  {2013})}\BibitemShut {NoStop}%
\bibitem [{\citenamefont {Tezuka}\ and\ \citenamefont
  {Garcia-Garcia}(2010)}]{Tezuka2010}%
  \BibitemOpen
  \bibfield  {author} {\bibinfo {author} {\bibfnamefont {M.}~\bibnamefont
  {Tezuka}}\ and\ \bibinfo {author} {\bibfnamefont {A.~M.}\ \bibnamefont
  {Garcia-Garcia}},\ }\href {\doibase 10.1103/PhysRevA.82.043613} {\bibfield
  {journal} {\bibinfo  {journal} {Phys. Rev. A}\ }\textbf {\bibinfo {volume}
  {82}},\ \bibinfo {pages} {43613} (\bibinfo {year} {2010})}\BibitemShut
  {NoStop}%
\bibitem [{\citenamefont {Feigel'man}\ \emph {et~al.}(2007)\citenamefont
  {Feigel'man}, \citenamefont {Ioffe}, \citenamefont {Kravtsov}, \citenamefont
  {Yuzbashyan},\ and\ \citenamefont {Feigel'man}}]{Feigelman2007}%
  \BibitemOpen
  \bibfield  {author} {\bibinfo {author} {\bibfnamefont {M.~V.}\ \bibnamefont
  {Feigel'man}}, \bibinfo {author} {\bibfnamefont {L.~B.}\ \bibnamefont
  {Ioffe}}, \bibinfo {author} {\bibfnamefont {V.~E.}\ \bibnamefont {Kravtsov}},
  \bibinfo {author} {\bibfnamefont {E.~A.}\ \bibnamefont {Yuzbashyan}}, \ and\
  \bibinfo {author} {\bibfnamefont {M.~V.}\ \bibnamefont {Feigel'man}},\
  }\href@noop {} {\bibfield  {journal} {\bibinfo  {journal} {Phys. Rev. Lett.}\
  }\textbf {\bibinfo {volume} {98}},\ \bibinfo {pages} {27001} (\bibinfo {year}
  {2007})}\BibitemShut {NoStop}%
\bibitem [{\citenamefont {Burmistrov}\ \emph {et~al.}(2012)\citenamefont
  {Burmistrov}, \citenamefont {Gornyi},\ and\ \citenamefont
  {Mirlin}}]{Burmistrov2012}%
  \BibitemOpen
  \bibfield  {author} {\bibinfo {author} {\bibfnamefont {I.~S.}\ \bibnamefont
  {Burmistrov}}, \bibinfo {author} {\bibfnamefont {I.~V.}\ \bibnamefont
  {Gornyi}}, \ and\ \bibinfo {author} {\bibfnamefont {A.~D.}\ \bibnamefont
  {Mirlin}},\ }\href {\doibase 10.1103/PhysRevLett.108.017002} {\bibfield
  {journal} {\bibinfo  {journal} {Phys. Rev. Lett.}\ }\textbf {\bibinfo
  {volume} {108}},\ \bibinfo {pages} {017002} (\bibinfo {year}
  {2012})}\BibitemShut {NoStop}%
\bibitem [{\citenamefont {Feigel'man}\ \emph {et~al.}(2010)\citenamefont
  {Feigel'man}, \citenamefont {Ioffe}, \citenamefont {Kravtsov},\ and\
  \citenamefont {Cuevas}}]{Feigelman2010}%
  \BibitemOpen
  \bibfield  {author} {\bibinfo {author} {\bibfnamefont {M.~V.}\ \bibnamefont
  {Feigel'man}}, \bibinfo {author} {\bibfnamefont {L.~B.}\ \bibnamefont
  {Ioffe}}, \bibinfo {author} {\bibfnamefont {V.~E.}\ \bibnamefont {Kravtsov}},
  \ and\ \bibinfo {author} {\bibfnamefont {E.}~\bibnamefont {Cuevas}},\ }\href
  {\doibase 10.1016/j.aop.2010.04.001} {\bibfield  {journal} {\bibinfo
  {journal} {Ann. Phys. (N. Y).}\ }\textbf {\bibinfo {volume} {325}},\ \bibinfo
  {pages} {1390} (\bibinfo {year} {2010})}\BibitemShut {NoStop}%
\bibitem [{\citenamefont {Wegner}(1980)}]{Wegner1980}%
  \BibitemOpen
  \bibfield  {author} {\bibinfo {author} {\bibfnamefont {F.}~\bibnamefont
  {Wegner}},\ }\href {\doibase 10.1007/BF01325284} {\bibfield  {journal}
  {\bibinfo  {journal} {Zeitschrift f\"{u}r Phys. B Condens. Matter}\ }\textbf
  {\bibinfo {volume} {36}},\ \bibinfo {pages} {209} (\bibinfo {year}
  {1980})}\BibitemShut {NoStop}%
\bibitem [{\citenamefont {Castellani}\ and\ \citenamefont
  {Peliti}(1986)}]{Castellani1986}%
  \BibitemOpen
  \bibfield  {author} {\bibinfo {author} {\bibfnamefont {C.}~\bibnamefont
  {Castellani}}\ and\ \bibinfo {author} {\bibfnamefont {L.}~\bibnamefont
  {Peliti}},\ }\href {\doibase http://dx.doi.org/10.1088/0305-4470/19/8/004}
  {\bibfield  {journal} {\bibinfo  {journal} {J. Phys. A. Math. Gen.}\ }\textbf
  {\bibinfo {volume} {19}},\ \bibinfo {pages} {L429} (\bibinfo {year}
  {1986})}\BibitemShut {NoStop}%
\bibitem [{\citenamefont {Fal'ko}\ and\ \citenamefont
  {Efetov}(1995)}]{Falko1995}%
  \BibitemOpen
  \bibfield  {author} {\bibinfo {author} {\bibfnamefont {V.~I.}\ \bibnamefont
  {Fal'ko}}\ and\ \bibinfo {author} {\bibfnamefont {K.~B.}\ \bibnamefont
  {Efetov}},\ }\href@noop {} {\bibfield  {journal} {\bibinfo  {journal} {EPL
  (Europhysics Lett.}\ }\textbf {\bibinfo {volume} {32}},\ \bibinfo {pages}
  {627} (\bibinfo {year} {1995})}\BibitemShut {NoStop}%
\bibitem [{\citenamefont {Fyodorov}\ and\ \citenamefont
  {Mirlin}(1997)}]{Fyodorov1997}%
  \BibitemOpen
  \bibfield  {author} {\bibinfo {author} {\bibfnamefont {Y.~V.}\ \bibnamefont
  {Fyodorov}}\ and\ \bibinfo {author} {\bibfnamefont {A.~D.}\ \bibnamefont
  {Mirlin}},\ }\href {\doibase 10.1103/PhysRevB.55.R16001} {\bibfield
  {journal} {\bibinfo  {journal} {Phys. Rev. B}\ }\textbf {\bibinfo {volume}
  {55}},\ \bibinfo {pages} {R16001} (\bibinfo {year} {1997})}\BibitemShut
  {NoStop}%
\bibitem [{\citenamefont {Hikami}\ \emph {et~al.}(1980)\citenamefont {Hikami},
  \citenamefont {Larkin},\ and\ \citenamefont {Nagaoka}}]{Hikami1980}%
  \BibitemOpen
  \bibfield  {author} {\bibinfo {author} {\bibfnamefont {S.}~\bibnamefont
  {Hikami}}, \bibinfo {author} {\bibfnamefont {A.~I.}\ \bibnamefont {Larkin}},
  \ and\ \bibinfo {author} {\bibfnamefont {Y.}~\bibnamefont {Nagaoka}},\
  }\href@noop {} {\bibfield  {journal} {\bibinfo  {journal} {Prog. Theor.
  Phys.}\ }\textbf {\bibinfo {volume} {63}},\ \bibinfo {pages} {707} (\bibinfo
  {year} {1980})}\BibitemShut {NoStop}%
\bibitem [{\citenamefont {Lobos}\ \emph {et~al.}(2013)\citenamefont {Lobos},
  \citenamefont {Tezuka},\ and\ \citenamefont {Garcia-Garcia}}]{Lobos2013}%
  \BibitemOpen
  \bibfield  {author} {\bibinfo {author} {\bibfnamefont {A.~M.}\ \bibnamefont
  {Lobos}}, \bibinfo {author} {\bibfnamefont {M.}~\bibnamefont {Tezuka}}, \
  and\ \bibinfo {author} {\bibfnamefont {A.~M.}\ \bibnamefont
  {Garcia-Garcia}},\ }\href {\doibase 10.1103/PhysRevB.88.134506} {\bibfield
  {journal} {\bibinfo  {journal} {Phys. Rev. B}\ }\textbf {\bibinfo {volume}
  {88}},\ \bibinfo {pages} {134506} (\bibinfo {year} {2013})}\BibitemShut
  {NoStop}%
\bibitem [{\citenamefont {Abeles}\ \emph {et~al.}(1966)\citenamefont {Abeles},
  \citenamefont {Cohen},\ and\ \citenamefont {Cullen}}]{Abeles1966}%
  \BibitemOpen
  \bibfield  {author} {\bibinfo {author} {\bibfnamefont {B.}~\bibnamefont
  {Abeles}}, \bibinfo {author} {\bibfnamefont {R.~W.}\ \bibnamefont {Cohen}}, \
  and\ \bibinfo {author} {\bibfnamefont {G.~W.}\ \bibnamefont {Cullen}},\
  }\href {\doibase 10.1103/PhysRevLett.17.632} {\bibfield  {journal} {\bibinfo
  {journal} {Phys. Rev. Lett.}\ }\textbf {\bibinfo {volume} {17}},\ \bibinfo
  {pages} {632} (\bibinfo {year} {1966})}\BibitemShut {NoStop}%
\bibitem [{\citenamefont {de~Gennes}(1966)}]{DeGennes1966}%
  \BibitemOpen
  \bibfield  {author} {\bibinfo {author} {\bibfnamefont {P.}~\bibnamefont
  {de~Gennes}},\ }\href@noop {} {\emph {\bibinfo {title} {{Superconductivity of
  Metals and Alloys}}}}\ (\bibinfo  {publisher} {W.A. Bebjamin, inc.},\
  \bibinfo {address} {New York},\ \bibinfo {year} {1966})\BibitemShut {NoStop}%
\bibitem [{\citenamefont {Shanenko}\ \emph {et~al.}(2008)\citenamefont
  {Shanenko}, \citenamefont {Croitoru},\ and\ \citenamefont
  {Peeters}}]{Shanenko2008}%
  \BibitemOpen
  \bibfield  {author} {\bibinfo {author} {\bibfnamefont {A.}~\bibnamefont
  {Shanenko}}, \bibinfo {author} {\bibfnamefont {M.}~\bibnamefont {Croitoru}},
  \ and\ \bibinfo {author} {\bibfnamefont {F.}~\bibnamefont {Peeters}},\ }\href
  {\doibase 10.1103/PhysRevB.78.024505} {\bibfield  {journal} {\bibinfo
  {journal} {Phys. Rev. B}\ }\textbf {\bibinfo {volume} {78}},\ \bibinfo
  {pages} {024505} (\bibinfo {year} {2008})}\BibitemShut {NoStop}%
\bibitem [{\citenamefont {Ma}\ and\ \citenamefont {Lee}(1985)}]{Ma1985}%
  \BibitemOpen
  \bibfield  {author} {\bibinfo {author} {\bibfnamefont {M.}~\bibnamefont
  {Ma}}\ and\ \bibinfo {author} {\bibfnamefont {P.}~\bibnamefont {Lee}},\
  }\href {\doibase 10.1103/PhysRevB.32.5658} {\bibfield  {journal} {\bibinfo
  {journal} {Phys. Rev. B}\ }\textbf {\bibinfo {volume} {32}},\ \bibinfo
  {pages} {5658} (\bibinfo {year} {1985})}\BibitemShut {NoStop}%
\bibitem [{\citenamefont {Efetov}(1983)}]{efetov1983supersymmetry}%
  \BibitemOpen
  \bibfield  {author} {\bibinfo {author} {\bibfnamefont {K.}~\bibnamefont
  {Efetov}},\ }\href@noop {} {\bibfield  {journal} {\bibinfo  {journal}
  {Advances in Physics}\ }\textbf {\bibinfo {volume} {32}},\ \bibinfo {pages}
  {53} (\bibinfo {year} {1983})}\BibitemShut {NoStop}%
\bibitem [{\citenamefont {Mirlin}(2000)}]{Mirlin2000}%
  \BibitemOpen
  \bibfield  {author} {\bibinfo {author} {\bibfnamefont {A.}~\bibnamefont
  {Mirlin}},\ }\href {\doibase 10.1016/S0370-1573(99)00091-5} {\bibfield
  {journal} {\bibinfo  {journal} {Phys. Rep.}\ }\textbf {\bibinfo {volume}
  {326}},\ \bibinfo {pages} {259} (\bibinfo {year} {2000})}\BibitemShut
  {NoStop}%
\bibitem [{\citenamefont {Evers}\ and\ \citenamefont
  {Mirlin}(2008)}]{Evers2008}%
  \BibitemOpen
  \bibfield  {author} {\bibinfo {author} {\bibfnamefont {F.}~\bibnamefont
  {Evers}}\ and\ \bibinfo {author} {\bibfnamefont {A.}~\bibnamefont {Mirlin}},\
  }\href {\doibase 10.1103/RevModPhys.80.1355} {\bibfield  {journal} {\bibinfo
  {journal} {Rev. Mod. Phys.}\ }\textbf {\bibinfo {volume} {80}},\ \bibinfo
  {pages} {1355} (\bibinfo {year} {2008})}\BibitemShut {NoStop}%
\bibitem [{\citenamefont {Sac\'{e}p\'{e}}\ \emph {et~al.}(2008)\citenamefont
  {Sac\'{e}p\'{e}}, \citenamefont {Chapelier}, \citenamefont {Baturina},
  \citenamefont {Vinokur}, \citenamefont {Baklanov},\ and\ \citenamefont
  {Sanquer}}]{Sacepe2008}%
  \BibitemOpen
  \bibfield  {author} {\bibinfo {author} {\bibfnamefont {B.}~\bibnamefont
  {Sac\'{e}p\'{e}}}, \bibinfo {author} {\bibfnamefont {C.}~\bibnamefont
  {Chapelier}}, \bibinfo {author} {\bibfnamefont {T.}~\bibnamefont {Baturina}},
  \bibinfo {author} {\bibfnamefont {V.}~\bibnamefont {Vinokur}}, \bibinfo
  {author} {\bibfnamefont {M.}~\bibnamefont {Baklanov}}, \ and\ \bibinfo
  {author} {\bibfnamefont {M.}~\bibnamefont {Sanquer}},\ }\href {\doibase
  10.1103/PhysRevLett.101.157006} {\bibfield  {journal} {\bibinfo  {journal}
  {Phys. Rev. Lett.}\ }\textbf {\bibinfo {volume} {101}},\ \bibinfo {pages}
  {157006} (\bibinfo {year} {2008})}\BibitemShut {NoStop}%
\bibitem [{\citenamefont {Mayoh}\ and\ \citenamefont
  {Garc\'{\i}a-Garc\'{\i}a}(2014)}]{Mayoh2014a}%
  \BibitemOpen
  \bibfield  {author} {\bibinfo {author} {\bibfnamefont {J.}~\bibnamefont
  {Mayoh}}\ and\ \bibinfo {author} {\bibfnamefont {A.~M.}\ \bibnamefont
  {Garc\'{\i}a-Garc\'{\i}a}},\ }\href {\doibase 10.1103/PhysRevB.90.134513}
  {\bibfield  {journal} {\bibinfo  {journal} {Phys. Rev. B}\ }\textbf {\bibinfo
  {volume} {90}},\ \bibinfo {pages} {134513} (\bibinfo {year}
  {2014})}\BibitemShut {NoStop}%
\bibitem [{\citenamefont {Dubi}\ \emph {et~al.}(2007)\citenamefont {Dubi},
  \citenamefont {Meir},\ and\ \citenamefont {Avishai}}]{Dubi2007}%
  \BibitemOpen
  \bibfield  {author} {\bibinfo {author} {\bibfnamefont {Y.}~\bibnamefont
  {Dubi}}, \bibinfo {author} {\bibfnamefont {Y.}~\bibnamefont {Meir}}, \ and\
  \bibinfo {author} {\bibfnamefont {Y.}~\bibnamefont {Avishai}},\ }\href
  {\doibase 10.1038/nature06180} {\bibfield  {journal} {\bibinfo  {journal}
  {Nature}\ }\textbf {\bibinfo {volume} {449}},\ \bibinfo {pages} {876}
  (\bibinfo {year} {2007})}\BibitemShut {NoStop}%
\bibitem [{\citenamefont {Ghosh}\ and\ \citenamefont
  {Mandal}(2013)}]{Ghosh2013}%
  \BibitemOpen
  \bibfield  {author} {\bibinfo {author} {\bibfnamefont {S.}~\bibnamefont
  {Ghosh}}\ and\ \bibinfo {author} {\bibfnamefont {S.~S.}\ \bibnamefont
  {Mandal}},\ }\href {\doibase 10.1103/PhysRevLett.111.207004} {\bibfield
  {journal} {\bibinfo  {journal} {Phys. Rev. Lett.}\ }\textbf {\bibinfo
  {volume} {111}},\ \bibinfo {pages} {207004} (\bibinfo {year}
  {2013})}\BibitemShut {NoStop}%
\bibitem [{\citenamefont {Quintanilla}\ and\ \citenamefont
  {Ziff}(2007)}]{Quintanilla2007}%
  \BibitemOpen
  \bibfield  {author} {\bibinfo {author} {\bibfnamefont {J.}~\bibnamefont
  {Quintanilla}}\ and\ \bibinfo {author} {\bibfnamefont {R.}~\bibnamefont
  {Ziff}},\ }\href {\doibase 10.1103/PhysRevE.76.051115} {\bibfield  {journal}
  {\bibinfo  {journal} {Phys. Rev. E}\ }\textbf {\bibinfo {volume} {76}},\
  \bibinfo {pages} {051115} (\bibinfo {year} {2007})}\BibitemShut {NoStop}%
\bibitem [{\citenamefont {Kapitulnik}\ and\ \citenamefont
  {Kotliar}(1985)}]{Kapitulnik1985}%
  \BibitemOpen
  \bibfield  {author} {\bibinfo {author} {\bibfnamefont {A.}~\bibnamefont
  {Kapitulnik}}\ and\ \bibinfo {author} {\bibfnamefont {G.}~\bibnamefont
  {Kotliar}},\ }\href {\doibase 10.1103/PhysRevLett.54.473} {\bibfield
  {journal} {\bibinfo  {journal} {Phys. Rev. Lett.}\ }\textbf {\bibinfo
  {volume} {54}},\ \bibinfo {pages} {473} (\bibinfo {year} {1985})}\BibitemShut
  {NoStop}%
\bibitem [{\citenamefont {Evers}\ and\ \citenamefont
  {Mirlin}(2000)}]{Evers2000}%
  \BibitemOpen
  \bibfield  {author} {\bibinfo {author} {\bibfnamefont {F.}~\bibnamefont
  {Evers}}\ and\ \bibinfo {author} {\bibfnamefont {A.}~\bibnamefont {Mirlin}},\
  }\href {\doibase 10.1103/PhysRevLett.84.3690} {\bibfield  {journal} {\bibinfo
   {journal} {Phys. Rev. Lett.}\ }\textbf {\bibinfo {volume} {84}},\ \bibinfo
  {pages} {3690} (\bibinfo {year} {2000})}\BibitemShut {NoStop}%
\end{thebibliography}%
 \appendix 
\section{The importance of the mean level spacing, $\delta_L$ on the matrix element}\label{Ap:dl}
\begin{figure}[h]
\begin{centering}
\includegraphics[width=\wid\textwidth]{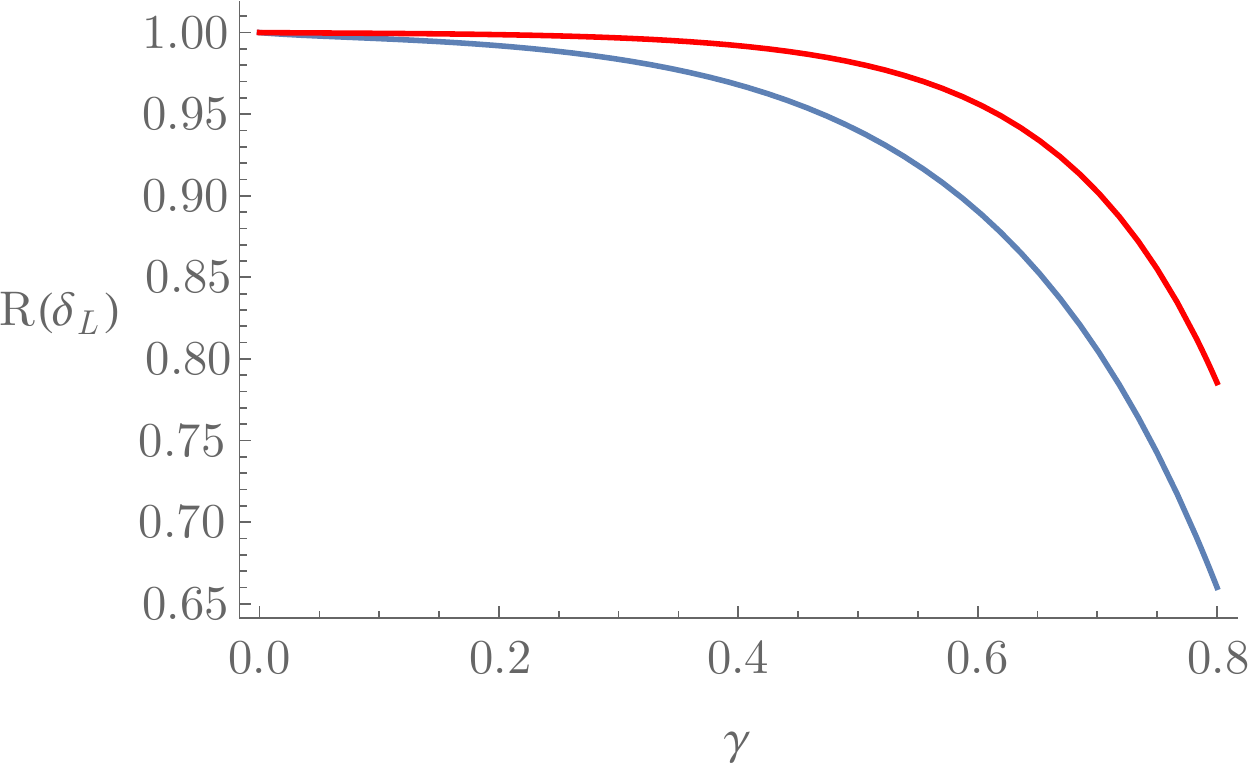}
\caption{Comparison of $R(\delta_L)$ for $\epsilon_D/\delta_L=100$(Blue) and $\epsilon_D/\delta_L=1000$(Red). Corresponding to the limit where BCS mean-field theory breaks-down, $\delta_L\sim\Delta_{\gamma=0}$, and the case for a clean metal, $\delta_L\ll\Delta_{\gamma=0}$, respectively. $R(\delta_L)$ is independent of $\epsilon_D/E_0$ and $\lambda$. We note there will be good agreement between results calculated using the simple matrix, Eq. (\ref{ME}), and results calculated with a careful treatment of the region around, $\delta_L$ Eq. (\ref{ME_dl}), when $\gamma\ll1$ and $\delta_L\ll\Delta_0$. }\label{F_R(d)}
\end{centering}
\end{figure}

In the work above it is assumed that the matrix element always follows Eq. (\ref{ME}) however the matrix element is known to saturate for states sufficiently close in energy. To see the effect of this saturation we can propose a matrix element which interpolates smoothly between these two behaviours,
\begin{equation}\label{ME_dl}
I(\epsilon,\epsilon')=\left(\frac{E_0}{\sqrt{(\epsilon-\epsilon')^2+\delta_L^2}}\right)^{\gamma}
\end{equation}
evaluating about the Fermi energy to zeroth order in $\gamma$,
\begin{equation}
1=\frac{\lambda}{2}\int_{-\epsilon_D}^{\epsilon_D} \frac{1}{\sqrt{\epsilon'^2+\Delta_\gamma^2}}\left(\frac{E_0}{\sqrt{\epsilon^2+\delta_L^2}}\right)^{\gamma} d\epsilon'
\end{equation}
\begin{equation}\label{cf2}
\frac{1}{\lambda}=\frac{\epsilon_DE_0^\gamma}{\Delta_\gamma \delta_L^\gamma}F_1\left(\frac{1}{2};\frac{1}{2},\frac{\gamma}{2};\frac{3}{2};-\frac{\epsilon_D^2}{\Delta_\gamma^2},-\frac{\epsilon_D^2}{\delta_L^2}\right)
\end{equation}
where $F_1$ is the Appell hypergeometric function. To compare the results of Eq. (\ref{cf1}) to the results of Eq. (\ref{cf2}) we define,
\begin{equation}
R(\delta_L)=\left(\frac{\epsilon_D}{\delta_L}\right)^{\gamma}\frac{(1-\gamma)F_1\left(\frac{1}{2};\frac{1}{2},\frac{\gamma}{2};\frac{3}{2};-\frac{\epsilon_D^2}{\Delta_\gamma^2},-\frac{\epsilon_D^2}{\delta_L^2}\right)}
{ \pFq{2}{1}\left({\frac{1}{2},\frac{1-\gamma}{2};\frac{3-\gamma}{2};-\frac{\epsilon_D^2}{\Delta_\gamma^2}}\right)}
\end{equation}
 Such that $R(\delta_L)\sim 1$ implies good agreement between the two forms of the matrix element and the role of $\delta_L$ may be neglected. We plot this function for different values of $\delta_L$ corresponding to, $\delta_L\sim\Delta_{\gamma=0}$ the point at which mean-field BCS treatment breaks down and $\delta_L\ll \Delta_{\gamma=0}$, which is the case for a bulk metal. The later case will hold for $\gamma\ll1$. We see that in both cases good agreement exists between the two forms of the matrix element up to moderate values of $\gamma$. see figure \ref{F_R(d)}.

\section{Energy dependence of the order parameter at zero temperature}\label{Ap_S2}
The energy dependence of the order parameter is obtained from the following generalized gap equation,
\begin{equation}\label{gap_eq_t01}
\Delta(\epsilon)=\frac{\lambda}{2}\int_{-\epsilon_D}^{\epsilon_D}\frac{\Delta(\epsilon')}{\sqrt{\epsilon'^2+\Delta^2(\epsilon')}}\left|\frac{E_0}{\epsilon-\epsilon'}\right|^\gamma d\epsilon'.
\end{equation}
where we assume that we are in the limit of weak multifractality such that $\gamma\ll1$. It is not in general acceptable to assume $\left(\frac{E_0}{|\epsilon|}\right)^\gamma$ is small as $E_0$ may be very large compared to $\epsilon$ as discussed in the introduction. For this reason we expand the matrix elements as,
\begin{equation}\begin{split}
I(\epsilon,\epsilon')=&\left|\frac{E_0}{\epsilon'}\right|^\gamma e^{-\gamma\ln\left|1-\frac{\epsilon}{\epsilon'}\right|}\\
=&\left|\frac{E_0}{\epsilon'}\right|^\gamma \left( 1-\gamma\ln\left|1-\frac{\epsilon}{\epsilon'}\right| +\mathcal{O}(\gamma^2)\right) 
\end{split}
\end{equation}
the logarithmic terms resulting from this expansion are acceptable as under integration they result in small corrections and so the series is convergent in $\gamma$.
We can also expand the left-most parts of the gap equation in powers of $\gamma$ using the ansatz,
\begin{equation}\label{gapenergy1}
\Delta(\epsilon)=\Delta_\gamma(1+\gamma f_1(\epsilon)+\gamma^2f_2(\epsilon) +\ldots)
\end{equation}

For example, to first order in $\gamma$,
\begin{widetext}
\begin{equation}\label{gap_exp}
1+\gamma f_1(\epsilon)+\mathcal{O}(\gamma^2)=\frac{\lambda}{2}\int_{-\epsilon_D}^{\epsilon_D}\left( \frac{1}{(\epsilon'^2+\Delta_\gamma^2)^{1/2}}+\gamma\frac{\epsilon'^2 f_1(\epsilon')}{(\epsilon'^2+\Delta_\gamma^2)^{3/2}} +\mathcal{O}(\gamma^2) \right) \left|\frac{E_0}{\epsilon'}\right|^\gamma 
\left( 1-\gamma\ln\left|1-\frac{\epsilon}{\epsilon'}\right| +\mathcal{O}(\gamma^2) \right) d\epsilon'
\end{equation}
\end{widetext}
The gap equation can now be solved for $\Delta_\gamma,f_1,f_2$, and higher terms if necessary, by collecting terms according to their $\gamma$ dependence.
\subsection{Zeroth order approximation}
Collecting the terms of order $\left|\frac{E_0}{\epsilon'}\right|^\gamma$ we find,
\begin{equation}
1=\frac{\lambda}{2}\int_{-\epsilon_D}^{\epsilon_D} \frac{1}{\sqrt{\epsilon'^2+\Delta_\gamma^2}} \left|\frac{E_0}{\epsilon'}\right|^\gamma d\epsilon'
\end{equation}
Carrying out the integral, 
\begin{equation}\label{cf1}
\frac{1}{\lambda}=\frac{E_0^{\gamma}\epsilon_D^{1-\gamma}}{\Delta_\gamma(1-\gamma)}\pFq{2}{1}\left({\frac{1}{2},\frac{1-\gamma}{2};\frac{3-\gamma}{2};-\frac{\epsilon_D^2}{\Delta_\gamma^2}}\right)
\end{equation}
where $\pFq{2}{1}(a,b;c;d)$ is the hypergeometric function. We define $\Delta_\gamma$ as the solution to this equation which corresponds approximately to the spectroscopic gap, namely, the minimum energy excitation at the Fermi energy. In section \ref{sec:zero_beh} we will carry out a full analysis of $\Delta_\gamma$. For now we focus on determining the energy dependence of the gap $\Delta(\epsilon)$.
\subsection{First order approximation}
Collecting the terms of order $\gamma\left|\frac{E_0}{\epsilon'}\right|^\gamma$ from Eq. (\ref{gap_exp})
\begin{equation}\label{f1_wrk}
\begin{split}
& f_1(\epsilon)=\\ &
\frac{\lambda}{2}\int_{-\epsilon_D}^{\epsilon_D}\left[\frac{\epsilon'^2 f_1(\epsilon')}{(\epsilon'^2+\Delta_\gamma^2)^{3/2}} \left|\frac{E_0}{\epsilon'}\right|^\gamma -\frac{ \ln\left|1-\frac{\epsilon}{\epsilon'}\right| }{\sqrt{\epsilon'^2+\Delta_\gamma^2}}\left|\frac{E_0}{\epsilon'}\right|^\gamma\right]\dd\epsilon'
\end{split}
\end{equation}
We solve Eq. (\ref{f1_wrk}) using the ansatz, $f_1(\epsilon)=h_1(\epsilon)+c_1$ 
 where $c_1$ is a constant and we define $h_1(\epsilon)$ as the closed function,
\begin{equation}
h_1(\epsilon)=-\frac{\lambda}{2}\int_{-\epsilon_D}^{\epsilon_D}\frac{ \ln\left|1-\frac{\epsilon}{\epsilon'}\right| }{\sqrt{\epsilon'^2+\Delta_\gamma^2}}\left|\frac{E_0}{\epsilon'}\right|^\gamma \dd\epsilon'
\end{equation}
After solving for $c_1$ we find that the leading correction to $\Delta_\gamma$ is given by,
\begin{equation}
f_1(\epsilon)=h_1(\epsilon)+\frac{\frac{\lambda}{2}\int_{-\epsilon_D}^{\epsilon_D}\frac{\epsilon'^2 h_1(\epsilon')}{(\epsilon'^2+\Delta_\gamma^2)^{3/2}} \left|\frac{E_0}{\epsilon'}\right|^\gamma \dd\epsilon'}{1-\frac{\lambda}{2}\int_{-\epsilon_D}^{\epsilon_D}\frac{\epsilon'^2}{(\epsilon'^2+\Delta_\gamma^2)^{3/2}} \left|\frac{E_0}{\epsilon'}\right|^\gamma \dd\epsilon'}
\end{equation}
\subsection{Second order approximation}
The treatment for the second order correction, $\gamma^2\left|\frac{E_0}{\epsilon'}\right|^\gamma$, is identical to the first order case. Using a similar ansatz we find, 
\begin{widetext}
\begin{equation}\label{f2}
 f_2(\epsilon)=h_2(\epsilon)+\frac{\frac{\lambda}{2}\int_{-\epsilon_D}^{\epsilon_D}\frac{\epsilon'^2 h_2(\epsilon')}{(\epsilon'^2+\Delta_\gamma^2)^{3/2}} \left|\frac{E_0}{\epsilon'}\right|^\gamma \dd\epsilon'}{1-\frac{\lambda}{2}\int_{-\epsilon_D}^{\epsilon_D}\frac{\epsilon'^2}{(\epsilon'^2+\Delta_\gamma^2)^{3/2}} \left|\frac{E_0}{\epsilon'}\right|^\gamma \dd\epsilon'}-\frac{3\lambda\Delta_\gamma^2}{4}\int_{-\epsilon_D}^{\epsilon_D}\frac{f_1(\epsilon')^2\epsilon'^2}{(\epsilon'^2+\Delta_\gamma^2)^{5/2}}\left|\frac{E_0}{\epsilon'}\right|^\gamma \dd\epsilon'
\end{equation}
where,
\begin{equation}
h_2(\epsilon)=\frac{\lambda}{2}\int_{-\epsilon_D}^{\epsilon_D}\left[\frac{\ln^2\left|1-\frac{\epsilon}{\epsilon'}\right|}{2\sqrt{\epsilon'^2+\Delta_\gamma^2}}\left|\frac{E_0}{\epsilon'}\right|^\gamma-\frac{\epsilon'^2\ln\left|1-\frac{\epsilon}{\epsilon'}\right| f_1(\epsilon')}{(\epsilon'^2+\Delta_\gamma^2)^{3/2}}\left|\frac{E_0}{\epsilon'}\right|^\gamma\right] \dd\epsilon'
\end{equation}
\end{widetext}

\section{Derivation of $T_{c\gamma}$}\label{Ap:tc}
Starting with,
\begin{equation}
1=\lambda\int_0^{\epsilon_D}\left(\frac{E_0}{\epsilon}\right)^\gamma\frac{\tanh(\beta_c\epsilon/2)}{\epsilon}\dd\epsilon
\end{equation}
let $x=\beta\epsilon/2$
\begin{equation}
1=\lambda \left(\frac{E_0\beta_c}{2}\right)^\gamma \int_0^{\frac{\beta_c\epsilon_D}{2}}\frac{\tanh(x)}{x^{1+\gamma}}\dd x
\end{equation}
We can carry out the integration by rewriting it as,
\begin{widetext}
\begin{equation}\label{ap1}
\begin{split}
 \int_0^{\frac{\beta_c\epsilon_D}{2}}\frac{\tanh(x)}{x^{1+\gamma}}\dd x
&=\int_0^1\frac{\tanh(x)}{x^{1+\gamma}}\dd x+\int_1^{\frac{\beta_c\epsilon_D}{2}}\left(\frac{1}{x^{1+\gamma}}-\frac{2}{x^{1+\gamma}( e^{2x}+1)}\right)\dd x\\
&=\frac{1}{\gamma}\left(1-\left(\frac{\beta_c\epsilon_D}{2}\right)^{-\gamma}\right)+\int_0^1\frac{\tanh(x)}{x^{1+\gamma}}\dd x-\int_1^{\frac{\beta_c\epsilon_D}{2}}\frac{2}{x^{1+\gamma}( e^{2x}+1)}\dd x
\end{split}
\end{equation}

Note the last line is only true if $\gamma\neq0$. We examine each of the remaining integrals in turn. 
\begin{equation}
\begin{split}
\int_0^1\frac{\tanh(x)}{x^{1+\gamma}}\dd x&=2\int_0^1\frac{\sinh(x)}{x^{1+\gamma}}(e^{-x}-e^{-2x}+e^{-5x}-\ldots)\dd x\\
\end{split}
\end{equation}
where we have used $\sech(x)=2(e^{-x}-e^{-3x}+e^{-5x}-\ldots)$. Integrating term by term and combining the results we find,
\begin{equation}\label{ap2}
\int_0^1\frac{\tanh(x)}{x^{1+\gamma}}\dd x=-\frac{1}{\gamma}+2^{\gamma+1}\Gamma(-\gamma)(1^\gamma-2^\gamma+3^\gamma-\ldots)
+2(\mathrm{E}_{1+\gamma}(2)-\mathrm{E}_{1+\gamma}(4)+\mathrm{E}_{1+\gamma}(6)-\ldots)
\end{equation}
\end{widetext}
where $\mathrm{E}_{n}(x)$ is the exponential integral function
\begin{equation}
\mathrm{E}_{n}(x)=\int_1^\infty\frac{e^{-xt}}{t^n}\dd t.
\end{equation}
Note the series $(1^\gamma-2^\gamma+3^\gamma-\ldots)$ is apparently not convergent. We know the integral is convergent and evaluate by taking the analytic continuation,
\begin{equation}\label{ap3}
(1^\gamma-2^\gamma+3^\gamma-\ldots)=(1-2^{\gamma+1})\zeta(-\gamma)
\end{equation}
where $\zeta(x)$ is the Riemann zeta function.

Now consider the integral,
\begin{equation}
\int_1^{\frac{\beta_c\epsilon_D}{2}}\frac{2}{x^{1+\gamma}( e^{2x}+1)}\dd x
\end{equation}
This function is well approximated ($k_BT_c\ll\epsilon_D$) by,
\begin{equation}\label{ap4}
\begin{split}
&\int_1^{\infty}\frac{2}{x^{1+\gamma}( e^{2x}+1)}\dd x=\int_1^{\infty}\frac{\sech(x)e^{-x}}{x^{1+\gamma}}\dd x\\
&=2(\mathrm{E}_{1+\gamma}(2)-\mathrm{E}_{1+\gamma}(4)+\mathrm{E}_{1+\gamma}(6)-\ldots)
\end{split}
\end{equation}
Combining eqs. (\ref{ap1}),(\ref{ap2}),(\ref{ap3}),(\ref{ap4}),
and rearranging gives the result,
\begin{equation}
k_BT_c=\epsilon_D C(\gamma)\left(\frac{1}{\lambda}\left(\frac{E_D}{E_0}\right)^\gamma+\frac{1}{\gamma}\right)^{-\frac{1}{\gamma}}
\end{equation}
\begin{equation}
C(\gamma)=\left[2(2^{\gamma+1}-1)\,\Gamma(-\gamma)\,\zeta(-\gamma)\right]^{\frac{1}{\gamma}}
\end{equation}
as required. 

\section{Analytical calculation of the spatial distribution of the order parameter}\label{Ap_S3}
We begin the calculation of the spatial distribution of the order parameter by computing the moments of $\Delta({\bf r})$ Eq. (\ref{twf0}),
\begin{equation}\label{dispar}
\langle\Delta^n({\bf r})\rangle=\int\dd{\bf r}\prod_{j=1}^n\left(\frac{\lambda V}{2}\int\frac{\Delta(\epsilon_{j})}{\sqrt{\Delta(\epsilon_{j})^2+\epsilon_{j}^2}}|\psi(\epsilon_{j},{\bf r})|^{2}d\epsilon_j \right) 
\end{equation}
where $\Delta(\epsilon_j)$ is given by Eq.(\ref{gapenergy}). 
 
It is clear that in order to proceed it is necessary to evaluate the following correlation function,
\begin{equation}
\tilde{P_q}=V^n \int\dd{\bf r}|\psi(\epsilon_{i_1},{\bf r})|^{2}|\psi(\epsilon_{i_2},{\bf r})|^{2}\ldots |\psi(\epsilon_{i_n},{\bf r})|^{2}.
\end{equation}
An exact analytical solution of Eq.(\ref{dispar}) is not possible however we shall see that by expanding in $\gamma \ll 1$ and keeping only the leading terms it is possible to find compact analytical solutions.

We assume without loss of generality that $\epsilon_{i_1}>\epsilon_{i_2}>\ldots>\epsilon_{i_n}$ and further always work in the case where $|\epsilon_{i_1}-\epsilon_{i_2}|\approx|\epsilon_{i_2}-\epsilon_{i_3}|\approx \ldots \approx|\epsilon_{i_{n-1}}-\epsilon_{i_n}|$. When the energy separation between the neighbouring eigenfunctions is small, $|\epsilon_{i_{k-1}}-\epsilon_{i_k}|\sim \delta_L$ we recover the results for the IPR,
\begin{equation}
\tilde{P_q}\sim L^{d_q(q-1)}
\end{equation}
whereas in the opposite limit $|\epsilon_{i_{k-1}}-\epsilon_{i_k}|\sim E_0$ the eigenfunctions become statistically independent and therefore,
\begin{equation}
\tilde{P_q}\approx V^{2n} \int\dd{\bf r}_1\ldots\int\dd{\bf r}_n|\psi(\epsilon_{i_1},{\bf r}_1)|^{2}\ldots |\psi(\epsilon_{i_n},{\bf r}_n)|^{2}\sim1
\end{equation}
Analogously to the derivation of Eq. (\ref{ME}), the scaling between these two limits can be approximated by,
\begin{equation}
\tilde{P_q}\sim\prod_{j=1}^{n-1}\left(\frac{E_0}{|\epsilon_{j}-\epsilon_{j+1}|}\right)^{\gamma_n}
\end{equation}
where $\gamma_n=1-\frac{d_n}{d}$.
The moments of the gap in real space can then be calculated from,
\begin{widetext}
\begin{equation}
\langle\Delta^n({\bf r})\rangle= \frac{\lambda}{2}\int d\epsilon_n \frac{\Delta(\epsilon_{n})}{\sqrt{\Delta(\epsilon_{n})^2+\epsilon_{n}^2}} \left(\prod_{j=1}^{n-1}\frac{\lambda}{2}\int d\epsilon_j\frac{\Delta(\epsilon_{j})}{\sqrt{\Delta(\epsilon_{j})^2+\epsilon_{j}^2}}\left(\frac{E_0}{|\epsilon_{j}-\epsilon_{j+1}|}\right)^{\gamma_n}\right)
\end{equation}
As when we solved the gap equation we expand in $\gamma$.
We consider the lowest order in $\gamma$ using, $\Delta(\epsilon)=\Delta_\gamma$,
\begin{equation}
\langle\Delta^n({\bf r})\rangle=\left(\frac{\lambda}{2}\right)^n\left(\prod_{j=1}^{n-1}\int d\epsilon_j \frac{\Delta_\gamma}{\sqrt{\Delta_\gamma^2+\epsilon_{j}^2}}\left(\frac{E_0}{|\epsilon_j|}\right)^{\gamma_n}\right)\int d\epsilon_n \frac{\Delta(\epsilon_{n})}{\sqrt{\Delta(\epsilon_{n})^2+\epsilon_{n}^2}}
\end{equation}
\end{widetext}
Carrying out the integrals,
and applying Eq.(\ref{gap_sg})
we find,
\begin{equation}
\langle\Delta^n({\bf r})\rangle=
\left(\Delta_\gamma\right)^{n}
\left(\frac{\epsilon_D}{E_0}\right)^{(\gamma-\gamma_n)(n-1)+\gamma}
\end{equation}
As was discussed in the introduction for a wide range of different systems, for example disorder in $d=2+\epsilon$ dimensions, it has been shown that the fractal dimension behaves like $d_n=d(1-\kappa n)$\cite{Falko1995,Mirlin2000,Evers2000}, where $\kappa^{-1}$ is proportional to the dimensionless conductance in the material. This dependence on $n$ applies for all $n$ less than some critical value $n_c$. For the systems we are interested in, this critical value is sufficiently large that the shape of the distribution will be well described by considering $d_n=d(1-\kappa n)$ for all $n$, as modifications to this value only affect very high order moments of the distribution.

Applying this result we can write our moments in the normalised form,
\begin{equation}
\frac{\langle\Delta^n({\bf r})\rangle}{\left(\Delta_\gamma\right)^{n}}
=e^{\kappa\ln(\epsilon_D/E_0) (3n-n^2)}
\end{equation}
from which it is trivial to write down the characteristic function associated with the distribution of $\Delta({\bf r})/\Delta_\gamma$,
\begin{equation}
\phi(t)=\sum_{n=0}^{\infty}\frac{(it)^n}{n!}e^{\kappa\ln(\epsilon_D/E_0)(3n-n^2)}
\end{equation}
By inspection, this is the characteristic function for a log-normal distribution,
\begin{equation}\label{PD1 }
\mathcal{P}\left(\frac{\Delta({\bf r})}{\Delta_\gamma}\right)=\frac{\Delta_\gamma}{\Delta({\bf r})\sqrt{2\pi}\sigma} \exp\left[-\frac{\left(\ln \left(\frac{\Delta({\bf r})}{\Delta_\gamma}\right)-\mu\right)^2}{2\sigma^2}\right]
\end{equation}
with $\mu=3\kappa\ln(\epsilon_D/E_0)$, $\sigma=\sqrt{2\kappa\ln(E_0/\epsilon_D)}$. The mean value for the distribution is,
\begin{equation}
\left\langle\frac{\Delta({\bf r})}{\Delta_\gamma}\right\rangle = \left(\frac{\epsilon_D}{E_0}\right)^{2\kappa}
\end{equation}
 and the variance is given by
\begin{equation}
\mathrm{Var}\left(\frac{\Delta({\bf r})}{\Delta_\gamma}\right)=\left(\frac{\epsilon_D}{E_0}\right)^{2\kappa}\left(1-\left(\frac{\epsilon_D}{E_0}\right)^{2\kappa}\right)
\end{equation}
\section{Solving the gap equation numerically}\label{Ap:num}

In principle solving the integral equation (\ref{gap_eq_t0}) is a difficult computational problem. We have developed a simple inexpensive algorithm to do this.

We first define an array of $n=200$ points $\epsilon_j$ equally spaced between $-\epsilon_D$ and $\epsilon_D$. We also define the gap at each of these points $\Delta_{i=0}(\epsilon_j)$ initialised it with a constant value $\Delta_0$. We then define a function which makes the array of the gap into a continuous function, $\Delta_{i=0}(\epsilon)$ using a high order polynomial interpolation. The integration can then be carried out using a standard numerical integration algorithm. We calculate $\Delta_{i=1}(\epsilon_j)$ using,
\begin{equation}
\Delta_{i+1}(\epsilon_j)=\frac{\lambda}{2}\int_{-\infty}^{\infty}\frac{\Delta_i(\epsilon')}{\sqrt{\epsilon'^2+\Delta_i^2(\epsilon')}}\left(\frac{E_0}{|\epsilon_j-\epsilon'|}\right)^\gamma d\epsilon'
\end{equation}
Now we iterate using $\Delta_{i=1}(\epsilon_j)$ as the input to the interpolation step. After several iterations the results converge to the correct value of the gap. We test convergence by defining the relative error,
\begin{equation}
 \text{err}_{i}=\frac{\sum_j |\Delta_{i}(\epsilon_j)-\Delta_{i-1}(\epsilon_j)|}{n\Delta_0}
\end{equation}
and take convergence to have been reached when $ \text{err}_{i}<10^{-6}$.


\end{document}